%% file: d7_double.tex
\documentclass[10pt,twocolumn,twoside]{IEEEtran}

\usepackage{cite}      

\usepackage{graphicx}  

\usepackage{subfigure} 

\usepackage{amsmath}   

\usepackage{amssymb}

\usepackage{epsfig,color}

\usepackage{theorem}

\usepackage{algorithm,algorithmic}

\def\vtor#1{\underline{#1}}
\newcommand{\ud}{\,\text{d}}

\def\TheTitle{Sparse image reconstruction for molecular imaging}

\makeatletter

\def\FigTwoSize{2in}
\def\FigThreeASize{1.5in}
\def\FigThreeBSize{1.6in}
\def\FigThreeCSize{1.5in}
\def\FigFourSize{1.6in}

\def\FigSixSize{1.6in}

\def\FigNineSize{1.6in}
\def\FigTwelveSize{1.6in}
\makeatother

\theoremstyle{plain}
\newtheorem{theorem}{Theorem}
\newtheorem{prop}{Proposition}
\newtheorem{definition}{Definition}

\begin{document}
%
\title{\TheTitle}
\author{Michael~Ting$^\ast$,~\IEEEmembership{Member,~IEEE,} 
	Raviv~Raich,~\IEEEmembership{Member,~IEEE,}
	and~Alfred~O.~Hero~III~\IEEEmembership{Fellow,~IEEE}%
\thanks{This research was partially supported by the Army Research Office under contract W911NF-05-1-0403.}%
\thanks{M.~Ting is with Seagate Technology, Pittsburgh, PA, 15222. Email: m\underline{ }ting@ieee.org; Phone: \mbox{+1 412 519 6946}; Fax: \mbox{+1 412 918 7010}.}%
\thanks{R.~Raich is with the Oregon State University, Corvallis, OR, 97331. Email: raich@eecs.oregonstate.edu; Phone: \mbox{+1 541 737 9862}; Fax: \mbox{+1 541 737 1300}.}%
\thanks{A.~O.~Hero~III is with the University of Michigan, Ann Arbor, MI, 48109. Email: hero@eecs.umich.edu; Phone: \mbox{+1 734 763 0564}; Fax: \mbox{+1 734 763 8041}.}}
\markboth{Submission to the {IEEE} Trans.~Image Processing}{Ting, Raich, and Hero:~\TheTitle}
%



\maketitle

\begin{abstract}
\input{abstract}
\end{abstract}


%
\IEEEpeerreviewmaketitle

\newcounter{ListCount}

\section{Introduction}
\label{sec:intro}

\input{intro_img}

\section{Problem formulation}
\label{sec:prob_form}
\input{prob_form}


\section{Algorithms} \label{sec:algo}

	\subsection{Bernoulli-Laplacian MAP/ML sparse estimators}
	\label{sec:map}
	\input{map}

	\subsection{Hybrid thresholding rule in the iterative framework}
	\label{sec:hybrid_iter}
	\input{hybrid_iter}

	\subsection{Using SURE to empirically estimate the hyperparameters}
	\label{sec:sure}
	\input{sure}

\section{Simulation study}
\label{sec:sim_study}
\input{sim_img}

\section{Summary and future directions}
\label{sec:summary}
\input{summary}

\bibliographystyle{IEEEtran}
\bibliography{IEEEabrv,mike_full}

\appendices

%
%

\section{Proofs of Section IV}
\label{app:proof_iv}
\input{proof_iv}

\section{Proofs of Section V}
\label{app:proof_v}
\input{proof_v}

\end{document}

%% file: abstract.tex
The application that motivates this paper is molecular imaging at the atomic level. When discretized at sub-atomic distances, the volume is inherently sparse. Noiseless measurements from an imaging technology can be modeled by convolution of the image with the system point spread function (psf). Such is the case with magnetic resonance force microscopy (MRFM), an emerging technology where imaging of an individual tobacco mosaic virus was recently demonstrated with nanometer resolution. We also consider additive white Gaussian noise (AWGN) in the measurements.
Many prior works of sparse estimators have focused on the case when $\mathbf{H}$ has low coherence; however, the system matrix $\mathbf{H}$ in our application is the convolution matrix for the system psf. A typical convolution matrix has high coherence. The paper therefore does not assume a low coherence $\mathbf{H}$. 
A discrete-continuous form of the Laplacian and atom at zero (LAZE) p.d.f.~used by Johnstone and Silverman is formulated, and two sparse estimators derived by maximizing the joint p.d.f.~of the observation and image conditioned on the hyperparameters. A thresholding rule that generalizes the hard and soft thresholding rule appears in the course of the derivation. This so-called hybrid thresholding rule, when used in the iterative thresholding framework, gives rise to the hybrid estimator, a generalization of the lasso. Unbiased estimates of the hyperparameters for the lasso and hybrid estimator are obtained via Stein's unbiased risk estimate (SURE). A numerical study with a Gaussian psf and two sparse images shows that the hybrid estimator outperforms the lasso.

%% file: intro_img.tex

The structures of biological molecules like proteins and viri are of interest to the medical community~\cite{modis04}. Existing methods for imaging at the nanometer or even sub-nanometer scale include atomic force microscopy (AFM), electron microscopy (EM), and X-ray crystallography~\cite{muller99,kuznetsov01}. At the sub-atomic scale, a molecule is naturally a sparse image. That is, the volume imaged consists of mostly space with a few locations occupied by atoms.
The application in particular that motivates this paper is MRFM~\cite{rugar04nature}, a technology that potentially offers advantages not existent in currently used methods. In particular, MRFM is non-destructive and capable of 3-d imaging. Recently, imaging of a biological sample with nanometer resolution was demonstrated~\cite{degen08_submitted}.
Given that MRFM and indeed even AFM~\cite{markiewicz95} measures the convolution of the image with a point spread function (psf), a deconvolution must be performed in order to obtain the molecular image.   
This paper considers the following problem: suppose one observes a linear transformation of a sparse image corrupted by AWGN. With only knowledge of the linear transformation and noise variance, the goal is to reconstruct the unknown sparse image.


The \emph{system matrix} $\mathbf{H}$ is the linear transformation that, in the case of MRFM, represents convolution with the MRFM psf. Several prior works are only applicable when the system matrix has small pairwise correlation, i.e., low coherence or low collinearity~\cite{tropp04,fuchs05,donoho06,tropp06}.
Others assume that the columns of $\mathbf{H}$ come from a specific random distribution, e.g., the uniform spherical ensemble (USE), or the uniform random projection ensemble (URPE)~\cite{donoho06_stomp}.
These assumptions are inapplicable when $\mathbf{H}$ represents convolution with the MRFM psf. In general, a convolution matrix for a continuous psf would not have low coherence. Such is the case with MRFM. The coherence of the simulated MRFM psf used in the simulation study section is at least 0.557.
%


The lasso, the estimator formed by maximizing the penalized likelihood criterion with a $l_1$ penalty on the image values~\cite{tibshirani96}, is known to promote sparsity in the estimate. The Bayesian interpretation of the lasso is the maximum a posteriori (MAP) estimate with an i.i.d.~Laplacian p.d.f.~on the image values~\cite{alliney94}. Consider the following: given $M$ i.i.d.~samples of a Laplacian distribution, the expected number of samples equal to $0$ is zero. The Laplacian p.d.f.~is more convincingly described as a heavy-tailed distribution rather than a sparse distribution. 
Indeed, when used in a suitable hierarchical model such as in sparse Bayesian learning~\cite{wipf04}, the Gaussian r.v., not commonly considered as a sparse distribution, results in a sparse estimator. While using a sparse prior is clearly not a necessary condition for formulating a sparse estimator, one wonders if a better sparse estimator can be formed if a sparse prior is used instead.


In~\cite{johnstone04}, the mixture of a Dirac delta and a symmetric, unimodal density with heavy tails is considered; a sparse denoising estimator is then obtained via marginal maximum likelihood (MML). The LAZE distribution is a specific member of the mixture family. Going through the same thought experiment previously mentioned with the LAZE distribution, one obtains an intuitive result: $Mw$ samples equal $0$, where $w$ is the weight placed on the Dirac delta. Unlike the Laplacian p.d.f., the LAZE p.d.f.~is both heavy-tailed and sparse. Under certain conditions, the estimator achieves the asymptotic minimax risk to within a constant factor~\cite[Thm.~1]{johnstone04}. 
The lasso estimator can be implemented in an iterative thresholding framework using the soft thresholding rule~\cite{figueiredo03,daubechies04}. Use of a thresholding rule based on the LAZE prior in the iterative thresholding framework can potentially result in better performance.


This paper develops several methods to enable Bayes-optimal nanoscale molecular imaging. In particular, advances are made in these three areas.
%
%
\begin{enumerate}
\item First, we introduce a mixed discrete-continuous LAZE prior for use in the MAP/maximum likelihood (ML) framework. Knowing only that the image is sparse, but lacking any precise information on the sparsity level, selection of the hyperparameters or regularization parameters has to be empirical or data-driven. The sparse image and hyperparameters are jointly estimated by maximizing the joint p.d.f.~of the observation and unknown sparse image conditioned on the hyperparameters. Two sparse Bernoulli-Laplacian MAP/ML estimators based on the discrete-continuous LAZE p.d.f.~are introduced: MAP1 and MAP2. 

\item The second contribution of the paper is the introduction of the hybrid estimator, which is formed by exclusively using the \emph{hybrid} thresholding rule in the iterative thresholding framework. The hybrid thresholding rule is a generalization of the soft and hard thresholding rules. In order to apply this to the molecular imaging problem, it is necessary to estimate the hyperparameters in a data-driven fashion.

\item Thirdly, SURE is applied to estimate the hyperparameter of lasso and of the hybrid estimator proposed above. The SURE-equipped versions of lasso and hybrid estimator are referred to as lasso-SURE and H-SURE. Our lasso-SURE result is a generalization of the results in~\cite{donoho95, ng99}. 
Alternative lasso hyperparameter selection methods exist, e.g.,~\cite{yuan05}. In~\cite{yuan05}, however, a prior is placed on the support of the image values that discourages the selection of high correlated columns of $\mathbf{H}$. Since the $\mathbf{H}$ we consider has columns that are highly correlated, this predisposes a certain amount of separation between the support of the estimated image values , i.e., the sparse image estimate will be resolution limited.
A number of other general-purpose techniques exist as well, e.g., cross validation (CV), generalized CV (GCV), MML~\cite{thompson91}. Some are, however, more tractable than others. For example, a closed form expression of the marginal likelihood cannot be obtained for the Laplacian prior: approximations have to be made~\cite{alliney94}. 
\end{enumerate}


A simulation study is performed. In the first part, LS, oracular LS, SBL, stagewise orthogonal matching pursuit (StOMP), and the four proposed sparse estimators, are compared. Two image types (one binary-valued and another based on the LAZE p.d.f.) are studied under two signal-to-noise ratio (SNR) conditions (low and high).
MAP2 has the best performance in the two low SNR cases. In one of the high SNR cases, H-SURE has the best performance, while in the other, SBL is arguably the best performing method. When the hyperparameters are estimated via SURE, H-SURE is sparser than lasso-SURE and achieves lower $l_p$ error for $p=0,1,2$ as well as lower detection error $E_d$.
In the second part of the numerical study, the performance of the proposed sparse estimators is studied across the range of SNRs between the low and high values considered in the first part. 
A 3-d reconstruction example is given in the third part, where the LS and lasso-SURE estimator are compared. This serves to demonstrate the applicability of lasso-SURE on a relatively large problem.


The paper is organized into the following sections. First, the sparse image deconvolution problem is formulated in Section~\ref{sec:prob_form}. 
The algorithms are discussed in Section~\ref{sec:algo}: there are three parts to this section.
The two MAP/ML estimators based on the discrete-continuous LAZE prior are derived in Section~\ref{sec:map}. This is followed by the introduction of the hybrid estimator in Section~\ref{sec:hybrid_iter}. Stein's unbiased risk estimate is applied in Section~\ref{sec:sure} to derive lasso-SURE and H-SURE. 
Section~\ref{sec:sim_study} contains a numerical study comparing the proposed algorithms with several existing sparse reconstruction methods. A summary of the work and future directions in Section~\ref{sec:summary} concludes the paper.

%% file: prob_form.tex
%
%

Consider a 2-d or 3-d image, and denote its vector version by $\vtor \theta \in \mathbb{R}^M$. In this paper, $\vtor\theta$ is assumed to be \emph{sparse}, viz., the percentage of non-zero $\theta_i$ is small. Suppose that the measurement $\vtor y \in \mathbb{R}^N$ is given by
%
%
\begin{equation}
\vtor y = \mathbf{H}\vtor\theta + \vtor w, \text{ where } 
	\vtor w \sim \mathcal{N}(\vtor 0,\sigma^2\mathbf{I}),
	\label{eqn:gaussian:obs_familiar}
\end{equation}
where $\mathbf{H} \in \mathbb{R}^{N \times M}$ is termed the \emph{system matrix}, and $\vtor w$ is AWGN. The problem considered can be stated as: given $\vtor y$, $\mathbf{H}$, and $\sigma>0$, estimate $\vtor\theta$ knowing that it is sparse. 
Without loss of generality, one can assume that the columns of $\mathbf{H}$ have unit $l_2$ norm. In the problem formulation, note that knowledge of the sparseness of $\vtor\theta$, viz., $\|\vtor\theta\|_0$, is \emph{not} known a priori.

It should be noted that, while the sparsity considered in (\ref{eqn:gaussian:obs_familiar}) is in the natural basis of $\vtor\theta$, a wavelet basis has been considered in other works, e.g.~\cite{ng99}. 
It may be possible to re-formulate (\ref{eqn:gaussian:obs_familiar}) using some other basis so that the corresponding system matrix has low coherence. This question is beyond the scope of the paper.
The emphasis here is on (\ref{eqn:gaussian:obs_familiar}) and on sparsity in the natural basis.
If $\mathbf{H}$ had full column rank, an equivalent problem formulation is available. Since $(\mathbf{H}^\prime\mathbf{H})$ is invertible, (\ref{eqn:gaussian:obs_familiar}) can be re-written as
%
%
\begin{equation}
\vtor{\tilde{y}} = \vtor\theta + \vtor{\tilde{w}}, \text{ where } 
	\vtor{\tilde{w}} \sim 
		\mathcal{N}( \vtor 0,\sigma^2\mathbf{H}^\dagger(\mathbf{H}^\dagger)^\prime )
	\label{eqn:gaussian:obs_equiv}
\end{equation}
where $\vtor{\tilde{y}} \triangleq \mathbf{H}^\dagger \vtor y$; $\mathbf{H}^\dagger \triangleq (\mathbf{H}^\prime\mathbf{H})^{-1}\mathbf{H}^\prime$ is the pseudoinverse of $\mathbf{H}$; and $\vtor{\tilde{w}} \triangleq \mathbf{H}^\dagger \vtor w$ is colored Gaussian noise. Deconvolution of $\vtor\theta$ from $\vtor y$ in AWGN is therefore equivalent to denoising of $\vtor\theta$ in colored Gaussian noise. In the special case that $\mathbf{H}$ is orthonormal, $\tilde{\vtor w}$ is also AWGN.

%% file: map.tex

This section considers the case when the discrete-continuous i.i.d.~LAZE prior is used for $p(\vtor\theta|\vtor\zeta)$, with $\vtor\theta$ and $\vtor\zeta$ simultaneously estimated via MAP/ML. For the continuous distribution, $\vtor\theta,\vtor\zeta$ are obtained as the maximizers of the conditional density $p(\vtor y,\vtor\theta|\vtor\zeta)$, viz.,
%
%
\begin{equation} \label{eqn:gauss_map:theta_phi}
\hat{\vtor\theta}, \hat{\vtor\zeta} 
	= \mathop{\text{argmax}}_{\vtor\theta,\vtor\zeta} \; \log p(\vtor y,\vtor\theta|\vtor\zeta) 
\end{equation}
If $\vtor\zeta$ were constant, $\hat{\vtor\theta}$ obtained from (\ref{eqn:gauss_map:theta_phi}) would be the MAP estimate. If $\vtor\theta$ were constant, the resulting $\hat{\vtor\zeta}$ would be the ML estimate. Since these two principles are at work, it cannot be said that the estimates obtained via (\ref{eqn:gauss_map:theta_phi}) are strictly MAP or ML. 


Recall that the LAZE p.d.f.~is given by
%
%
\begin{equation} \label{eqn:laze}
p(\theta_i) = (1-w) \delta(\theta_i) + w \gamma(\theta_i;a),
\end{equation}
where $\gamma(x;a)=(1/2)ae^{-a|x|}$ is the Laplacian p.d.f.
The Dirac delta function is difficult to work with in the context of maximizing the conditional p.d.f.~in (\ref{eqn:gauss_map:theta_phi}). Consider then a mixed \emph{discrete-continuous} version of (\ref{eqn:laze}). Define the random variables $\tilde{\theta}_i$ and $I_i$ such that $\theta_i = I_i \tilde{\theta}_i$, $1 \le i \le M$. The r.v.s $\tilde{\theta}_i, I_i$ have the following density: 
%
%
\begin{align}
I_i & = \left\{ \begin{array}{cc}
	0 & \text{with probability } (1-w) \\
	1 & \text{with probability } w \end{array} \right.
	\label{eqn:gauss_map_prior:1} \\
p(\tilde{\theta}_i|I_i) & = \left\{ \begin{array}{cc}
	g(\tilde{\theta}_i) & I_i = 0 \\
	\gamma(\tilde{\theta}_i;a) & I_i = 1 \end{array} \right.
	\label{eqn:gauss_map_prior:2},
\end{align}
where $g(\cdot)$ is some p.d.f.~that will be specified later on. It is assumed that $(\tilde{\theta}_i, I_i)$ are i.i.d. $I_i$ assumes the role of the Dirac delta: its introduction necessitates use of the auxiliary density $g$ in (\ref{eqn:gauss_map_prior:2}). Instead of (\ref{eqn:gauss_map:theta_phi}), consider the optimality criterion
%
%
\begin{equation}
\hat{\tilde{\vtor\theta}},\hat{\vtor I},\hat{\vtor \zeta} 
	= \mathop{\text{argmax}}_{\tilde{\vtor\theta},\vtor I,\vtor \zeta} \; 
		\log p(\tilde{\vtor \theta},\vtor I|\vtor y, \vtor \zeta)
	\label{eqn:gauss_map:theta_I_phi}
\end{equation}
Let $\mathcal{I}_1 \triangleq \{i:I_i=1\}$ and $\mathcal{I}_0 \triangleq \overline{\mathcal{I}}_1 = \{i:I_i=0\}$. The maximization of (\ref{eqn:gauss_map:theta_I_phi}) is equivalent to the maximization of
%
%
\begin{align}
\Psi_\text{map} & \triangleq -\frac{\|\mathbf{H}\vtor\theta-\vtor y\|^2}{2\sigma^2}  
	+ (M-\|\vtor I\|_0)\log(1-w) \nonumber \\
&	+ \|\vtor I\|_0\log w + \sum_{i\in\mathcal{I}_1} \log\left(\frac{1}{2}ae^{-a|\tilde{\theta}_i|}\right) +
	\sum_{i\in\mathcal{I}_0} \log g(\tilde{\theta}_i)
	\label{eqn:gauss_map_theta_I_phi:2}
\end{align}
We propose to maximize (\ref{eqn:gauss_map_theta_I_phi:2}) in a block coordinate-wise fashion~\cite{fesslerbook} via Algorithm~\ref{alg:map}. Note that $\hat{\theta}_i=\hat{\tilde{\theta}}_i \hat{I}_i$. A superscript ``$(n)$'' attached to a variable indicates its value in the $n$th iteration.
%
%
\begin{algorithm}
\caption{Block coordinate maximization of MAP criterion $\Psi_\text{map}$.}
\label{alg:map}
\begin{algorithmic}[1]
\REQUIRE $\hat{\tilde{\vtor \theta}}^{(0)}$, $\hat{\vtor I}^{(0)}$, $\epsilon>0$
\STATE $n \leftarrow 0$
\REPEAT 
	\STATE $n \leftarrow n + 1$
	\STATE $\hat{\vtor\zeta}^{(n)} \leftarrow \text{argmax}_{\vtor\zeta} \, \Psi_\text{map}\left(\hat{\tilde{\vtor\theta}}^{(n-1)},\hat{\vtor I}^{(n-1)},\vtor\zeta \right)$ 
	\STATE $\hat{\tilde{\vtor\theta}}^{(n)},\hat{\vtor I}^{(n)} \leftarrow \text{argmax}_{\tilde{\vtor\theta},\vtor I} \, \Psi_\text{map}\left(\tilde{\vtor\theta},\vtor I,\hat{\vtor\zeta}^{(n)}\right)$ 
\UNTIL{ $\| \hat{\vtor\theta}^{(n)} - \hat{\vtor\theta}^{(n-1)} \| < \epsilon$ } 
\end{algorithmic}
\end{algorithm}

The p.d.f.~$g$ arises as an extra degree of freedom due to the introduction of the indicator variables $I_i$. Consider two cases: first, let $g(x)=\gamma(x;a)$ in (\ref{eqn:gauss_map_theta_I_phi:2}). This will give rise to the algorithm MAP1. Second, let $g(x)$ be an arbitrary p.d.f.~such that: (1) $|g(x)|<\infty$ for all $x \in \mathbb{R}$; (2) $\sup g(x)$ is attained for some $x \in \mathbb{R}$; and (3) $g(x)$ is \emph{independent} of $a,w$. By selecting $g$ that satisfies these three properties, the algorithm MAP2 is thus obtained.

\subsubsection{MAP1}


Let $\Psi_\text{map1}(\tilde{\vtor\theta},\vtor I,\vtor\zeta)$ denote the function obtained by setting $g(x)=\gamma(x;a)$.
%
%
%
%
%
%
Step (4) of Algorithm~\ref{alg:map} is determined by the solution to $\nabla_{\vtor\zeta} \Psi_\text{map1}=0$. This is solved as
%
%
\begin{equation}
\hat{a} = \frac{M}{\|\hat{\tilde{\vtor\theta}}\|_1} \text{ and }
\hat{w} = \frac{\|\hat{\vtor I}\|_0}{M}.
	\label{eqn:map1:stepi}
\end{equation}
It can be verified that the Hessian $\nabla_{\vtor\zeta}\nabla_{\vtor\zeta}^T \Psi_\text{map1}$ is negative definite for all $a>0$ and $0<w<1$.
Given $n$ samples $x_1,\ldots,x_n$ drawn from a Laplacian p.d.f.~$\gamma(\cdot;a)$, the ML estimate of $a$ is $\hat{a}_\text{ML} = n(\sum_{i=1}^n |x_i|)^{-1}$. The estimate $\hat{a}$ in (\ref{eqn:map1:stepi}) is therefore the ML estimate of $a$ where all of the $\hat{\tilde{\theta}}_i$s are used.

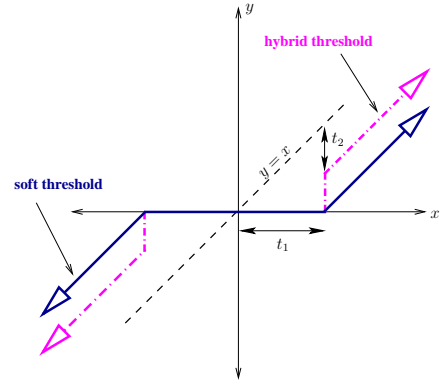
\begin{figure}[!h]
\centering
	\resizebox{!}{\FigTwoSize}{\input{hybrid_thres.pstex_t}}
	\caption{Hybrid thresholding rule.}
	\label{fig:hybrid_thres}
\end{figure}

The maximization in step (5) of Algorithm~\ref{alg:map} can be obtained by applying the EM algorithm~\cite{figueiredo03}. Recall that EM can be applied using $\vtor z = \vtor \theta + \alpha \vtor w_1$ as the \emph{complete data}, where $\vtor w_1 \sim \mathcal{N}(0,\alpha^2\mathbf{I})$ and $\alpha \le \sigma/\|\mathbf{H}\|_2$. Denote by $\vtor{\hat{\theta}}^{(n)}, \hat{\vtor{z}}^{(n)}$ the estimates in the $n$th EM iteration.
The E-step is the Landweber iteration
%
%
\begin{equation} \label{eqn:gaussian:em_1}
\hat{\vtor z}^{(n)} 
	= \hat{\vtor\theta}^{(n-1)} + \left(\frac{\alpha}{\sigma}\right)^2 \mathbf{H}^T ( \vtor y - \mathbf{H}\hat{\vtor\theta}^{(n-1)} ).
\end{equation}
%
%
%
%
%
%
%
%
%
%
%
%
Define the \emph{hybrid thresholding rule} as 
%
%
\begin{equation}
T_\textit{hy}(x;t_1,t_2) \triangleq (x-\text{sgn}(x)t_2)I(|x|>t_1).
\label{eqn:hybrid_thres},
\end{equation}
where $t_1$ and $t_2$ are restricted to $0 \le t_2 \le t_1$. See Fig.~\ref{fig:hybrid_thres}. This is a generalization of the soft and hard thresholding rules. The soft thresholding rule $T_s(x;t)=T_\textit{hy}(x;t,t)$, and the hard thresholding rule $T_h(x;t)=xI(|x|>t)=T_\textit{hy}(x;t,0)$. 
%
%
The M-step of the EM algorithm is given by
%
%
\begin{equation}
\theta_i = \left\{ \begin{array}{ll}
	T_\textit{hy}(\hat{z}_i^{(n)};a\alpha^2 + \kappa_1(\alpha,w), a\alpha^2) &
	0 < w \le \frac{1}{2} \\
	T_\text{s}(\hat{z}_i^{(n)};a\alpha^2) &
	\frac{1}{2} < w \le 1
	\end{array} \right.
	\label{eqn:map1:theta}
\end{equation}
where $\kappa_1(\alpha,w) \triangleq \sqrt{2\alpha^2\log((1-w)/w)}$. Recall that $\theta_i=\tilde{\theta}_i I_i$. If $w>1/2$, the soft-thresholding rule is applied in the $Q$-step of the EM iterations of MAP1. These iterations produce the lasso estimate with hyperparameter $\zeta=2a\alpha^2$. However, if $0 < w \le 1/2$, a larger thresholding value is used that increases the smaller $w$ becomes.

\subsubsection{MAP2}


From (\ref{eqn:gauss_map_prior:2}) and the assumptions on $g$, $\tilde{\theta}_i \neq 0$ w.p.~1. Consequently, the set
%
%
\begin{equation}
\mathcal{I}_1 = \{i:I_i=1\}=\{i:\theta_i \neq 0\} \; \text{w.p.~1}. 
\label{eqn:map2:assumption}
\end{equation}
This implies $\|\vtor I\|_0 = \|\vtor\theta\|_0$ w.p.~1. Apply (\ref{eqn:map2:assumption}) to the criterion to maximize, viz., (\ref{eqn:gauss_map_theta_I_phi:2}), and denote the result by $\Psi_\text{map2}(\tilde{\vtor\theta},\vtor I,\vtor\zeta)$. One gets
%
%
\begin{align}
\Psi_\text{map2}
	& = -\frac{1}{2\sigma^2}\|\mathbf{H}\vtor\theta-\vtor y\|^2 
	+ (M-\|\vtor I\|_0)\log(1-w) \nonumber \\
 	& + \|\vtor I\|_0\log w  
	+ \|\vtor\theta\|_0 \log \frac{a}{2} - a \|\vtor\theta\|_1 + \sum_{\{i:I_i=0\}} \log g(\tilde{\theta}_i)
	\label{eqn:map2:criterion}.
\end{align}
%
%
The maximization in step (i) is obtained by solving for $\nabla_{\vtor\zeta} \Psi_\text{map2}=0$, which produces
%
%
\begin{equation}
\hat{a} = \frac{\|\hat{\vtor\theta}\|_0}{\|\hat{\vtor\theta}\|_1} \text{ and }
\hat{w} = \frac{\|\hat{\vtor\theta}\|_0}{M}.
\label{eqn:map2:stepi}
\end{equation}
As before, one can verify that the Hessian $\nabla_{\vtor\phi}\nabla_{\vtor\phi}^T \Psi_2$ is negative definite for all $a>0$ and $0<w<1$.
It is instructive to compare the hyperparameter estimates of MAP1 vs.~MAP2, i.e.,~(\ref{eqn:map1:stepi}) vs.~(\ref{eqn:map2:stepi}). The main difference lies in the estimation of $a$. Assuming that the estimates $\hat{\vtor I}$ and $\hat{\vtor\theta}$ obey (\ref{eqn:map2:assumption}), one can re-write the MAP2 estimate $\hat{a}=|\mathcal{I}_1|/\sum_{i\in\mathcal{I}_1} |\hat{\tilde{\theta}}_i|$. This is the ML estimate using only the $\hat{\tilde{\theta}}_i, i \in \mathcal{I}_1$, i.e., the non-zero voxels. 
On the other hand, the MAP1 estimate of $a$ can be written as
%
%
\begin{equation}
\hat{a} = \frac{|\mathcal{I}_1|+|\mathcal{I}_0|}{
	\sum_{i\in\mathcal{I}_1} |\hat{\tilde{\theta}}_i| +
	\sum_{i\in\mathcal{I}_0} |\hat{\tilde{\theta}}_i| }
	\label{eqn:map1:stepi:a}.
\end{equation}
%
%


As with MAP1, the maximization in step (5) of Algorithm~\ref{alg:map} can be obtained by applying the EM algorithm with the complete data $\vtor z = \vtor\theta + \alpha \vtor{w}_1$. 
The E-step is given by (\ref{eqn:gaussian:em_1}), which is the same as MAP1's E-step.
Define 
%
%
\begin{equation} \label{eqn:map2:defs}
g^\ast \triangleq \sup_x g(x) \;\; \text{and} \;\; 
r \triangleq \frac{g^\ast}{a/2}\frac{1-w}{w}.
\end{equation}
%
%
%
%
%
%
The resulting $\vtor\theta$ in the M-step is given by the following thresholding rule
%
%
\begin{equation}
\theta_i = \left\{ \begin{array}{ll}
	T_\textit{hy}(\hat{z}^{(n)}_i; a\alpha^2 + \kappa_2(\alpha,r), a\alpha^2 ) & r \ge 1 \\
	T_\text{s}(\hat{z}^{(n)}_i;a\alpha^2) & 0 \le r < 1 \\
	\end{array}\right.
	\label{eqn:map2:theta}
\end{equation}
where $\kappa_2(\alpha,r) \triangleq \sqrt{2\alpha^2 \log r}$, which is similar to the M-step of MAP1. 
%
Indeed, the M-step of MAP1 can be obtained by setting $g^\ast = a/2$. 
Just like in MAP1, the EM iterations of MAP2 produce a larger threshold the sparser the hyperparameter $w$ is. As well, if $a$ is smaller, $r$ increases. Since the variance of the Laplacian $\gamma(\cdot;a)$ is $2/a^2$, a smaller $a$ implies a larger variance of the Laplacian. Use of a larger threshold is therefore appropriate.

The tuning parameter $g^\ast$ can be regarded as an extra degree of freedom that arises due to $g$ being independent of $a,w$. The MAP2 M-step is a function of $g^\ast$, and a suitable value has to be selected. In contrast, MAP1 has no free tuning parameter(s). 
%

%% file: hybrid_thres.pstex_t
\begin{picture}(0,0)%
\includegraphics{hybrid_thres.pstex}%
\end{picture}%
\setlength{\unitlength}{3947sp}%
\begingroup\makeatletter\ifx\SetFigFont\undefined%
\gdef\SetFigFont#1#2#3#4#5{%
  \reset@font\fontsize{#1}{#2pt}%
  \fontfamily{#3}\fontseries{#4}\fontshape{#5}%
  \selectfont}%
\fi\endgroup%
\begin{picture}(5246,4567)(2576,-6800)
\put(5713,-5228){\makebox(0,0)[lb]{\smash{{\SetFigFont{12}{14.4}{\rmdefault}{\mddefault}{\updefault}{\color[rgb]{0,0,0}$t_1$}%
}}}}
\put(6369,-3966){\makebox(0,0)[lb]{\smash{{\SetFigFont{12}{14.4}{\rmdefault}{\mddefault}{\updefault}{\color[rgb]{0,0,0}$t_2$}%
}}}}
\put(7546,-4854){\makebox(0,0)[lb]{\smash{{\SetFigFont{12}{14.4}{\rmdefault}{\mddefault}{\updefault}{\color[rgb]{0,0,0}$x$}%
}}}}
\put(5337,-2389){\makebox(0,0)[lb]{\smash{{\SetFigFont{12}{14.4}{\rmdefault}{\mddefault}{\updefault}{\color[rgb]{0,0,0}$y$}%
}}}}
\put(5553,-4384){\rotatebox{45.0}{\makebox(0,0)[lb]{\smash{{\SetFigFont{12}{14.4}{\rmdefault}{\mddefault}{\updefault}{\color[rgb]{0,0,0}$y=x$}%
}}}}}
\end{picture}%

%% file: hybrid_iter.tex

Define the \emph{hybrid estimator} to be the estimator formed by using the hybrid thresholding rule (\ref{eqn:hybrid_thres}) in the iterative framework~\cite[(24)]{figueiredo03}, viz.,
%
%
\begin{equation} \label{eqn:em_iter:hybrid}
 \vtor{\hat{\theta}}^{(n+1)} = S_{T_\textit{hy},\vtor\zeta} \left( \vtor{\hat{\theta}}^{(n)} +
	(\alpha/\sigma)^2 \mathbf{H}^\prime ( \vtor y - \mathbf{H} \vtor{\hat{\theta}}^{(n)} \right).
\end{equation}
where $S_{T_\textit{hy};\vtor\zeta}(\vtor x) = \sum_i T_\textit{hy}(x_i;\vtor\zeta) e_i$ and $\vtor e_i \in \mathbb{R}^M, i=1,\ldots,M$ are the standard unit vectors.
Due to the hybrid thresholding rule being a generalization of the soft thresholding rule, the hybrid estimator potentially offers better performance than lasso. The cost function of the hybrid estimator is given in Prop.~\ref{prop:hybrid:cost}.

\begin{prop} \label{prop:hybrid:cost}
Consider the iterations (\ref{eqn:em_iter:hybrid}) when $\|\mathbf{H}\|_2<1$ and $\alpha=\sigma$. The iterations minimize the cost function
%
%
\begin{align}
\Psi_{\vtor\zeta,\textit{hy}}(\vtor\theta) & = \|\mathbf{H}\vtor\theta-\vtor y\|_2^2 + 
	\sum_i J_1(\theta_i) \nonumber \\
\text{where: } J_1(x) & = I(|x|< \zeta_1-\zeta_2)[ 
			-(x-\text{sgn}(x)\zeta_1)^2 + 2 \zeta_1 \zeta_2 ] 
	\nonumber \\
	& \quad + I(|x|\ge \zeta_1-\zeta_2)(2\zeta_2|x|+\zeta_2^2) \label{eqn:hybrid:cost}
\end{align}
%

\end{prop}

\textbf{Proof}. This is an application of Thm.~\ref{thm:general:cost} in Appendix~\ref{app:proof_iv}. 
When $\zeta_1=\zeta_2=\zeta$, $J_1(x)=2\zeta|x|+\zeta^2$, which gives rise to the lasso estimator, as expected. 
$\blacksquare$

%% file: sure.tex
In this section, SURE is applied to estimate the regularization parameter of lasso and the hybrid estimator. Consider the $l_2$ risk measure
%
%
\begin{equation} \label{eqn:risk_theta:sure}
 R(\vtor\theta,\vtor\zeta) = \frac{1}{N} E_{\vtor Y}\|\vtor{\hat\theta}-\vtor{\theta}\|_2^2
\end{equation}
for lasso. Since $\vtor\theta$ is not known, this risk cannot be computed; however, one can compute an \emph{unbiased} estimate of the risk~\cite{stein81}. Denote the unbiased estimate by $\hat{R}(\vtor\zeta)$: $\vtor\zeta$ can then be estimated as $\vtor{\hat\zeta} = \text{argmin}_{\vtor\zeta \in \Omega} \hat{R}(\vtor\zeta)$, where $\Omega$ is the set of valid values. When $\mathbf{H}=\mathbf{I}$, an expression for $\hat{R}(\vtor\zeta)$ is derived in~\cite[(11)]{donoho95}. When $\mathbf{H} \neq \mathbf{I}$, however, Stein's unbiased estimate~\cite{stein81} cannot be applied to evaluate (\ref{eqn:risk_theta:sure}). In~\cite{ng99}, the alternative $l_2$ risk
%
%
\begin{equation} \label{eqn:risk_Htheta:sure}
 R(\vtor\theta,\vtor\zeta) = \frac{1}{N} E_{\vtor Y}\|\mathbf{H}(\vtor{\hat\theta}-\vtor{\theta})\|_2^2
\end{equation}
is proposed instead. Equation (\ref{eqn:risk_Htheta:sure}) was evaluated for a diagonal $\mathbf{H}$ in~\cite{ng99}.

The first theorem in this section generalizes the result of~\cite{ng99} by developing $\hat{R}(\vtor\zeta)$ for arbitrary full column rank $\mathbf{H}$. The second theorem in this section derives (\ref{eqn:risk_Htheta:sure}) when $\vtor{\hat{\theta}}$ is the hybrid estimator. For this result, $\mathbf{H}$ is also an arbitrary full column matrix. If the convolution matrix can be approximated by 2d or 3d circular convolution, the full column rank assumption is equivalent to the 2d or 3d DFT of the psf having no spectral nulls. The proofs of the two theorems are given in Appendix~\ref{app:proof_v}.

\subsubsection{SURE for lasso}

%
%
\begin{theorem} \label{thm:sure1}
Assume that the columns of $\mathbf{H}$ are linearly independent, and $\hat{\vtor\theta}$ is the lasso estimator. The unbiased risk estimate (\ref{eqn:risk_Htheta:sure}) is
%
%
\begin{equation} \label{eqn:thm_sure1:1}
\hat{R}(\zeta) = \sigma^2 + \frac{1}{N} \|\vtor e\|_2^2 + \frac{2\sigma^2}{N}\|\vtor{\hat{\theta}}\|_0
\end{equation}
where $\vtor e = \vtor y - \mathbf{H}\vtor{\hat\theta}$ is the reconstruction error.
%
%
\end{theorem}

\vspace{0.4ex}


Since the hyperparameter $\zeta \ge 0$, it can be estimated via
%
%
\begin{equation} \label{eqn:thm_sure1:2}
 \hat{\zeta} = \text{argmin}_{\zeta \ge 0} \hat{R}(\zeta)
\end{equation}
where $\hat{R}(\zeta)$ is given in (\ref{eqn:thm_sure1:1}). LARS can be used to compute (\ref{eqn:thm_sure1:2}). Note that LARS requires the linear independence of the columns of $\mathbf{H}$. The estimator $\vtor{\hat\theta}_l(\vtor{\hat\zeta})$ with $\vtor{\hat\zeta}$ obtained via (\ref{eqn:thm_sure1:2}) will be referred to as lasso-SURE.

\subsubsection{SURE for the hybrid estimator}

\label{subsec:img:hybrid_sure}


Several definitions are in order first. 
%
%
\begin{definition} \label{def:perm_pq}
Suppose that $\hat{\vtor\theta} \in \mathbb{R}^M$ has $\|\hat{\vtor\theta}\|_0=M-r$. Denote the non-zero components of $\hat{\vtor\theta}$ by $x_i$, $1 \le i \le M-r$. The permutation matrix $\mathbf{P}(\vtor{\hat\theta}) \in \mathbf{R}^{M \times M}$ is said to order the zero and non-zero components of $\hat{\vtor\theta}$ if  $\mathbf{P}\text{diag}(\hat{\vtor\theta})\mathbf{P}^\prime=\text{diag}(0,\ldots,0,x_1,\ldots,x_{M-r})$.
\end{definition}
Note that $\mathbf{P}$ in the above definition is not unique. As $\mathbf{P}$ is a permutation matrix, it is orthogonal.
%
%
\begin{definition} \label{def:submatrix}
For a matrix $\mathbf{A}=(a_{ij}) \in \mathbb{R}^{q \times r}$, let $c_n$ be a non-zero sequence of length at most $q$ s.t.~$1 \le c_n \le q$. Similarly, let $d_n$ be non-zero sequence of length at most $r$ s.t.~$1 \le d_n \le r$. The submatrix $\mathbf{A}[c_n,d_n] = (\alpha_{ij})$ is such that $\alpha_{ij} = a_{c_i,d_j}$. 
\end{definition}
Define $\Delta_\zeta \triangleq \zeta_1 - \zeta_2$ and 
%
%
\begin{equation} \label{eqn:thm_sure2:1}
\mathbf{U}(\vtor\theta) \triangleq \left\{ \begin{array}{cc}
	\text{diag}[\text{rect}(\frac{\theta_1}{2\Delta_\zeta}),\ldots,
		    \text{rect}(\frac{\theta_M}{2\Delta_\zeta})] & \Delta_\zeta > 0 \\
	\mathbf{0} & \Delta_\zeta = 0
                                           \end{array}
	\right.
\end{equation}
where $\text{rect}(x) = 1, |x| \le 1/2$ and $0$ otherwise. Recall that $0 \le \zeta_2 \le \zeta_1$ by assumption, so $\Delta_\zeta \ge 0$. Let $\mathbf{G}(\mathbf{H}) \triangleq \mathbf{H}^\prime\mathbf{H}$ denote the Gram matrix of $\mathbf{H}$. For a given $\vtor{\hat\theta}$, set
%
%
\begin{align}
 \mathbf{C}_1(\vtor{\hat\theta}) & \triangleq (\mathbf{P}\mathbf{G}(\mathbf{H})\mathbf{P}^\prime)[r+1:M,r+1:M]
	\label{eqn:thm_sure2:2} \;\; \text{ and } \\
 \mathbf{C}_2(\vtor{\hat\theta}) & \triangleq -\frac{1}{2}(\mathbf{P}\mathbf{U}(\hat{\vtor\theta})\mathbf{P}^\prime)[r+1:M,r+1:M]
	\label{eqn:thm_sure2:3},
\end{align}
where $\mathbf{P}$ is a matrix that orders the zero and non-zero components of $\vtor{\hat\theta}$.

\begin{theorem} \label{thm:sure2}
Suppose that the columns of $\mathbf{H}$ are linearly independent and that $\mathbf{G}(\mathbf{H})$ does not have an eigenvalue of $1/2$. 
With $\vtor{\hat\theta}$ denoting the hybrid estimator, the unbiased risk estimate (\ref{eqn:risk_Htheta:sure}) is
%
%
\begin{equation} \label{eqn:thm_sure2:4}
\hat{R}(\vtor\zeta) = \sigma^2 + \frac{1}{N} \|\vtor e\|_2^2 + 	
	\frac{2\sigma^2}{N}\text{tr}(\mathbf{C}_1[\mathbf{C}_1+\mathbf{C}_2]^{-1})	
\end{equation}
where $\vtor e = \vtor y - \mathbf{H}\vtor{\hat\theta}$.
%
%
%
\end{theorem}

To evaluate (\ref{eqn:thm_sure2:4}) for a particular $\vtor{\hat\theta}$, one would have to construct the matrix $\mathbf{P}$; then, invert the $(M-r) \times (M-r)$ matrix $(\mathbf{C}_1+\mathbf{C}_2)$. If $\hat{\vtor\theta}$ is sparse, $(M-r)$ is small, and the inversion would not be computationally demanding. 
The optimum $\vtor\zeta$ is the $\vtor \zeta \in \{ (\zeta_1,\zeta_2): \zeta_1 \ge 0, \, 0 \le \zeta_2 \le \zeta_1 \}$ that minimizes $\hat{R}(\vtor\zeta)$. The corresponding $\hat{\vtor\theta}_\textit{hy}(\vtor\zeta)$ would be the output. This method will be referred to as Hybrid-SURE, or for short, H-SURE.


%% file: sim_img.tex

In Section~\ref{sec:sim_img1}, the following classes of methods are compared: (i) least-squares (LS) and oracular LS; (ii) the proposed sparse reconstruction methods; and (iii) other existent sparse methods, viz., SBL and StOMP.


The LS solution is implemented via the Landweber algorithm~\cite{byrne04}. It provides a ``worst-case'' bound for the $l_2$ error, i.e., $\|\vtor e\|_2$. Since the LS estimate does not take into account the sparsity of $\vtor\theta$, one would expect it to have worse performance than estimates that do.
In the oracular LS method, on the other hand, one knows the support of $\vtor\theta$, and regresses the measurement $\vtor y$ on the corresponding columns of $\mathbf{H}$~\cite{candes05}. The oracular LS estimate consequently provides a ``best-case'' bound for the $l_2$ error; however, the oracular LS estimate is unimplementable in reality, as it requires prior knowledge of the support of $\vtor\theta$.
The second class of methods includes the two MAP/ML variants, MAP1 and MAP2; in addition, lasso-SURE and H-SURE are also tested.
Finally, in order to benchmark the proposed methods to other sparse methods, SBL and StOMP are included in the simulation study. The Sparselab toolbox is used to obtain the StOMP estimate. The CFAR and CFDR approaches to threshold selection are applied~\cite{donoho06_stomp}. For CFAR selection, the per-iteration false alarm rate of $1/50$ is used. For CFDR selection, the discovery rate is set to $0.5$. Although a multitude of other sparse reconstruction methods exist, they are not included in the simulation study due to a lack of space.

\begin{figure}[!h]
\centering
\begin{minipage}[b]{.48\linewidth}
	\centering
	\centerline{\epsfig{figure=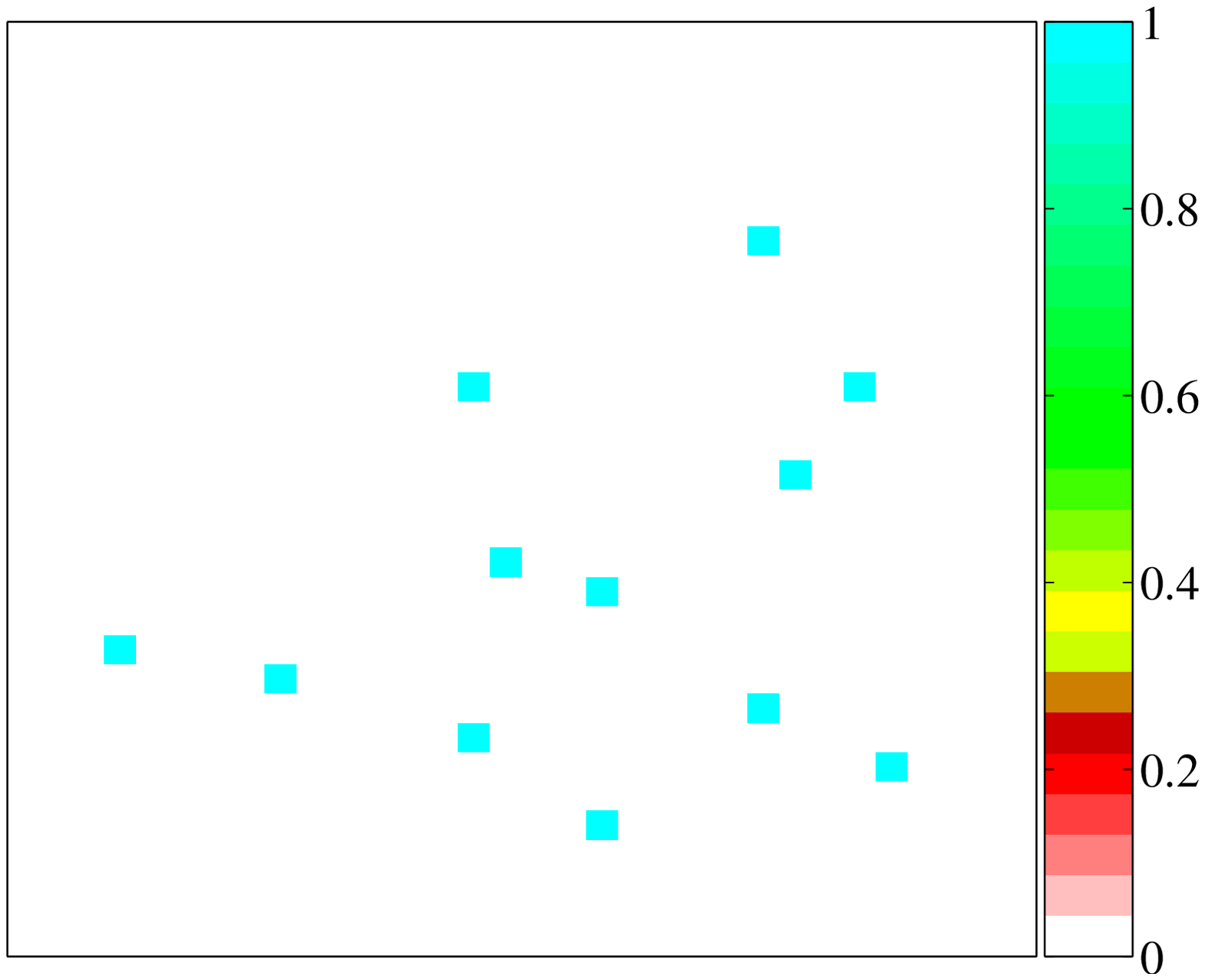,width=\FigThreeASize}}
	\vspace{0.25cm}
	\centerline{(a) Binary $\vtor\theta$}
\end{minipage} \hfill
\begin{minipage}[b]{0.48\linewidth}
	\centering
	\centerline{\epsfig{figure=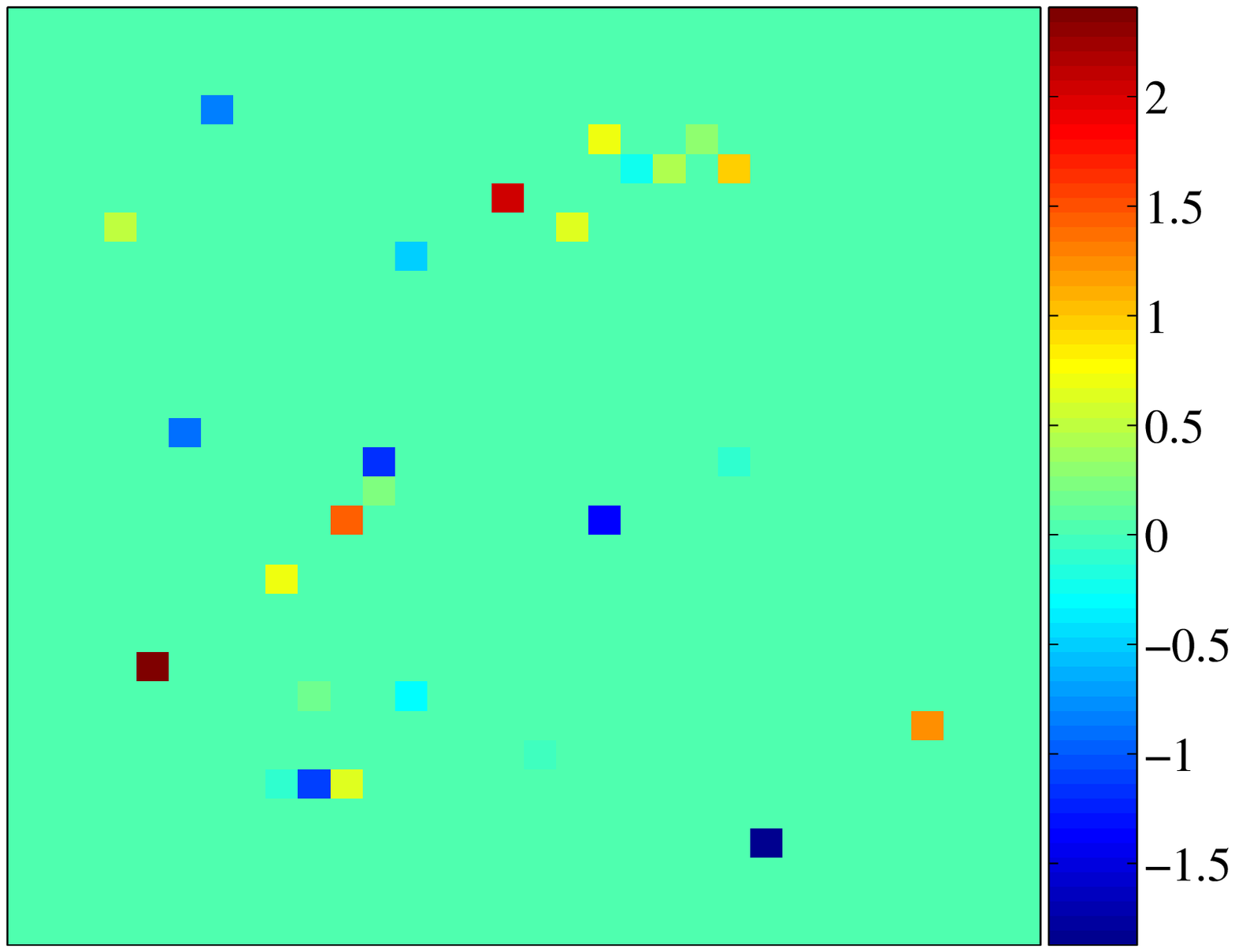,width=\FigThreeBSize}}
	\vspace{0.25cm}
	\centerline{(b) LAZE  $\vtor \theta$}
\end{minipage}

\vspace{0.25cm}

\begin{minipage}[b]{0.48\linewidth}
	\centering
	\centerline{\epsfig{figure=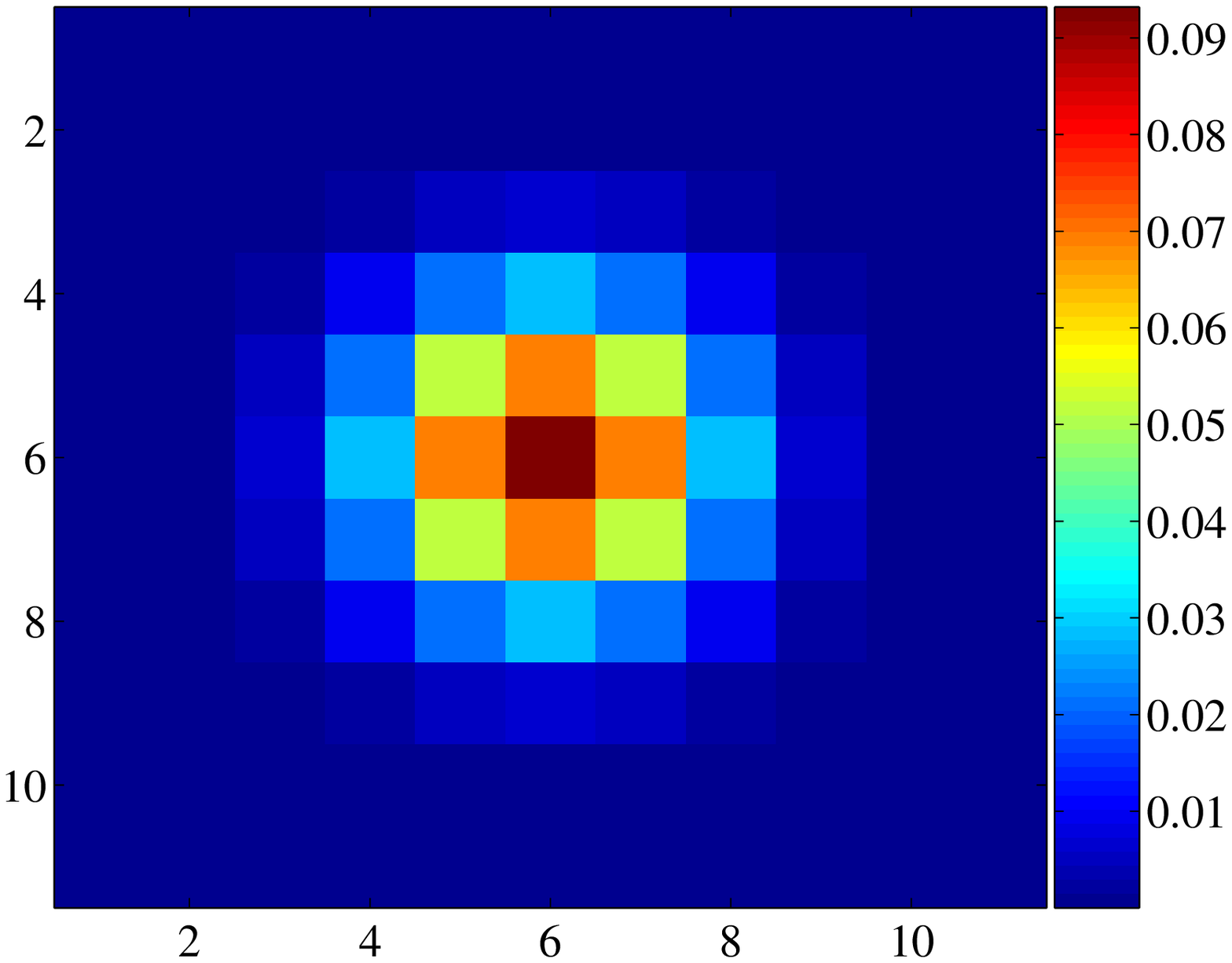,width=\FigThreeCSize}}
	\vspace{0.25cm}
	\centerline{(c) Gaussian blur psf}
\end{minipage}
\caption{Illustration of the two types of $\vtor\theta$ used in the simulations; as well, the Gaussian blur psf is shown.}
\label{fig:numerical_study}
\end{figure}

Two sparse images $\vtor\theta$ are investigated in Section~\ref{sec:sim_img1}: a binary-valued image, and an image based on the LAZE prior (\ref{eqn:laze}). The binary-valued image has 12 pixels set to one, and the rest are zero. The LAZE image, i.e., the image based on the LAZE prior, can be regarded as a realization of the LAZE prior with $a=1$ and $w=0.026$. They are depicted in Fig.~\ref{fig:numerical_study}a,b respectively. The two images are of size $32 \times 32$, as is $\vtor y$: so, $M=N=1024$. The matrix $\mathbf{H}$, of size $1024 \times 1024$, is the convolution matrix for the Gaussian blur point spread function (psf). In order to satisfy the requirements of Thm.~\ref{thm:sure1} and~\ref{thm:sure2}, the columns of $\mathbf{H}$ are linearly independent and $\mathbf{G}(\mathbf{H})$ does not have an eigenvalue of $1/2$. The Gaussian blur is illustrated in Fig.~\ref{fig:numerical_study}c.  
%
%

The Gaussian blur convolution matrix has columns that are highly correlated: the coherence $\mu=0.86089$. Let $\Lambda(\vtor\theta) \triangleq \{ i: \theta_i \neq 0\}$. The stability and support results of lasso all require that
%
%
\begin{equation} \label{eqn:erc_approx_general}
\mu |\Lambda(\vtor\theta)| \lessapprox c
\end{equation}
where $c=1/2$ or $1/4$ in order that some statement of recoverability holds~\cite{fuchs05,candes05,donoho06,tropp06}. For a given $\mathbf{H}$, (\ref{eqn:erc_approx_general}) places an upper bound on $|\Lambda(\vtor\theta)|$ for which recoverability of $\vtor\theta$ is assured in some fashion. With the Gaussian blur $\mathbf{H}$, $|\Lambda(\vtor\theta)| \lessapprox c/0.86089 < 1$ for both $c=1/4$ and $1/2$. Since $\|\vtor\theta\|_0=12$, the simulation study is outside of the coverage of existing recoverability theorems.


In Section~\ref{sec:sim_img2}, the performance of the proposed sparse methods over a range of SNRs is investigated. The binary-valued image and Gaussian blur psf are considered in this section. In addition to the proposed sparse methods, the LS estimate is included as a point of reference.
Lastly, a 3d MRFM example of dimension $128 \times 128 \times 32$ is given in Section~\ref{sec:sim_img4} comparing the LS estimate and lasso-SURE. This serves to illustrate the computational feasibility of lasso-SURE for a relatively large problem.


The proposed algorithms are implemented as previously outlined. The tuning parameter $g^\ast$ of MAP2 is set to $1/\sqrt{2}$ in Section~\ref{sec:sim_img1} and~\ref{sec:sim_img2}. 
LARS is used to compute the lasso-SURE estimator. H-SURE is suboptimally implemented: the minimizing $\vtor \zeta=(\zeta_1,\zeta_2)^\prime$ is obtained via two line searches. The first, along the $(1/\sqrt{2},1/\sqrt{2})$ direction in the $(\zeta_1, \zeta_2)$ plane, is done using lasso-SURE. A subsequent line search in the $(1,0)$ direction is performed, i.e., $\zeta_2$ is kept constant and $\zeta_1$ is increased. 
Define the SNR as $\text{SNR} \triangleq (N^{-1}\|\mathbf{H}\vtor\theta\|^2)/\sigma^2$, and the SNR in dB as $\text{SNR}_\text{dB} \triangleq 10 \log_{10} \text{SNR}$. 

\subsection{Error criteria}

Recall that the reconstruction error $\vtor e = \vtor\theta-\hat{\vtor\theta}$. Several error criteria are considered in the performance assessment of a sparse estimator.
%
%
\begin{itemize}

\item $\|\vtor e\|_p$ for $p=0,1,2$.

\item The \emph{detection error} criterion defined by
%
%
\begin{equation}
E_d(\vtor\theta,\hat{\vtor\theta};\delta) \triangleq \sum_{i=1}^M |I(\theta_i = 0) - I(|\hat{\theta}_i| < \delta)|
	\label{eqn:sim:1}
\end{equation}
Values of $\hat{\theta}_i$ such that $|\hat{\theta}_i|<\delta$ are considered equivalent to $0$. This is used to handle the effect of finite-precision computing. More importantly, it addresses the fact that, to the human observer, small non-zero values are not discernible from zero values. In the study, $\delta = 10^{-2} \|\vtor\theta\|_\infty$ is selected. This error criterion is effectively a 0-1 penalty on the support of $\vtor\theta$. Accurately determining the support of a sparse $\vtor\theta$ is more critical than its actual values~\cite{herrity06,tropp04}. 

\item The number of non-zero values of $\hat{\vtor\theta}$, i.e., \textbf{$\|\hat{\vtor\theta}\|_0$}. One would like $\|\hat{\vtor\theta}\|_0 \approx \|\vtor\theta\|_0$, which is small if $\vtor\theta$ is indeed sparse.

\end{itemize}


\newcommand{\best}[1]{\underline{#1}}

\subsection{Performance under low and high SNR}
\label{sec:sim_img1}
\input{sim_img1_tip}

\subsection{Performance vs.~SNR of the proposed reconstruction methods}
\label{sec:sim_img2}
\input{sim_img2_tip}


\subsection{MRFM reconstruction example}
\label{sec:sim_img4}
\input{sim_img4_tip}

%% file: sim_img1_tip.tex

The performance of the estimators is given in Table~\ref{tab:sim:binary} for the binary-valued $\vtor\theta$ with the $\text{SNR}$ equal to 1.76 dB (low SNR) and 20 dB (high SNR). 
The number reported in Table~\ref{tab:sim:binary} is the mean over the simulation runs. For each performance criterion, the best mean number is underlined. The oracular LS estimate is excluded from this assessment, as it cannot be implemented without prior knowledge. In terms of $\|\hat{\vtor\theta}\|_0$, the best number is the value closest to $\|\vtor\theta\|_0$. Recall that for the binary-valued image $\vtor\theta$, $\|\vtor\theta\|_0=12$. The best number for the other performance criterion is the value closest to $0$. 
%
%
%
%
\input{sim_img1_tip_tab1}



\begin{figure}[!h]
\centering

%
%
\begin{minipage}[b]{0.48\linewidth}
	\centering
	\centerline{\epsfig{figure=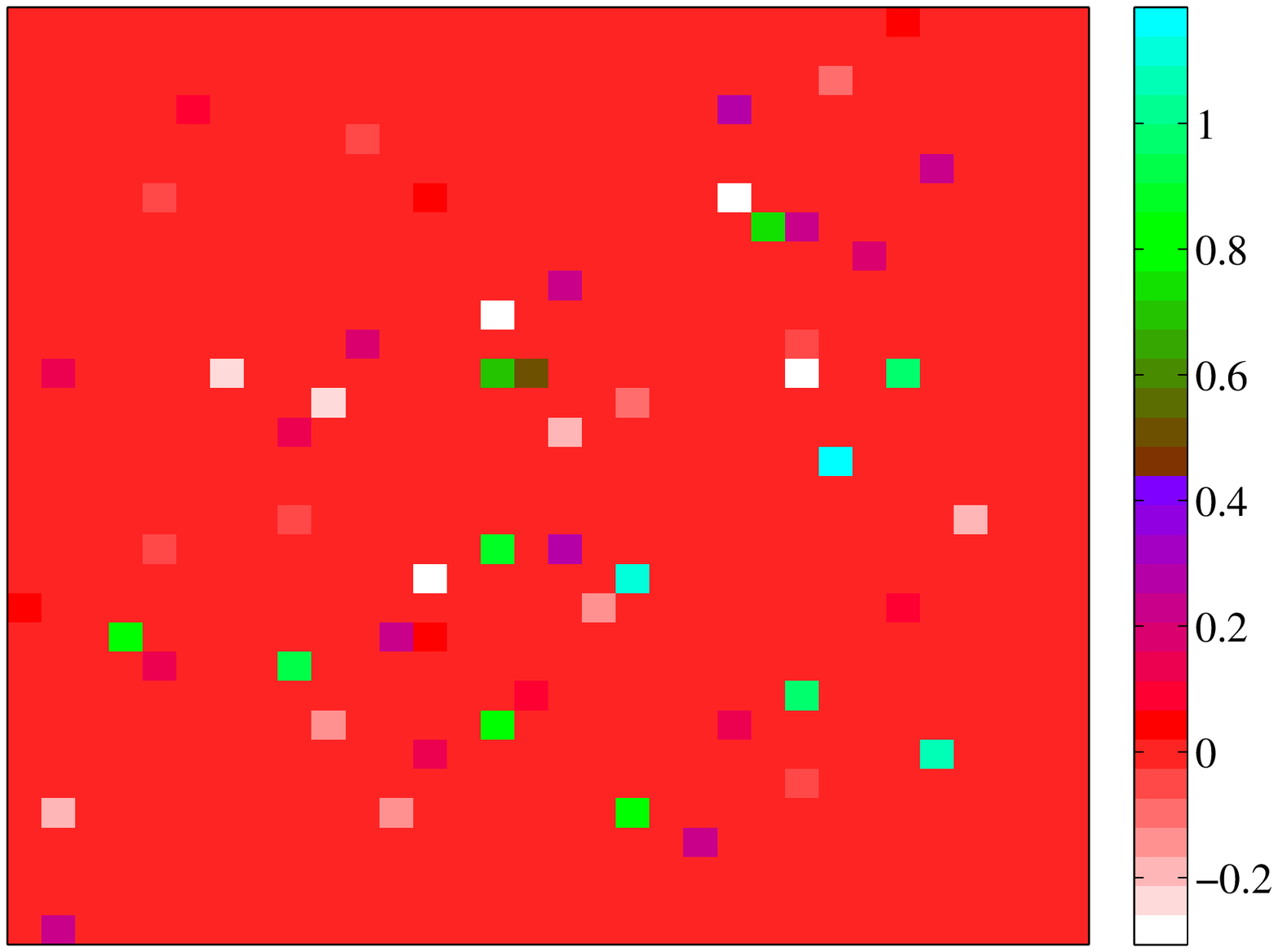,width=\FigFourSize}}
	\centerline{(a) SBL}
\end{minipage}
\hfill
%
%
\begin{minipage}[b]{0.48\linewidth}
	\centering
	\centerline{\epsfig{figure=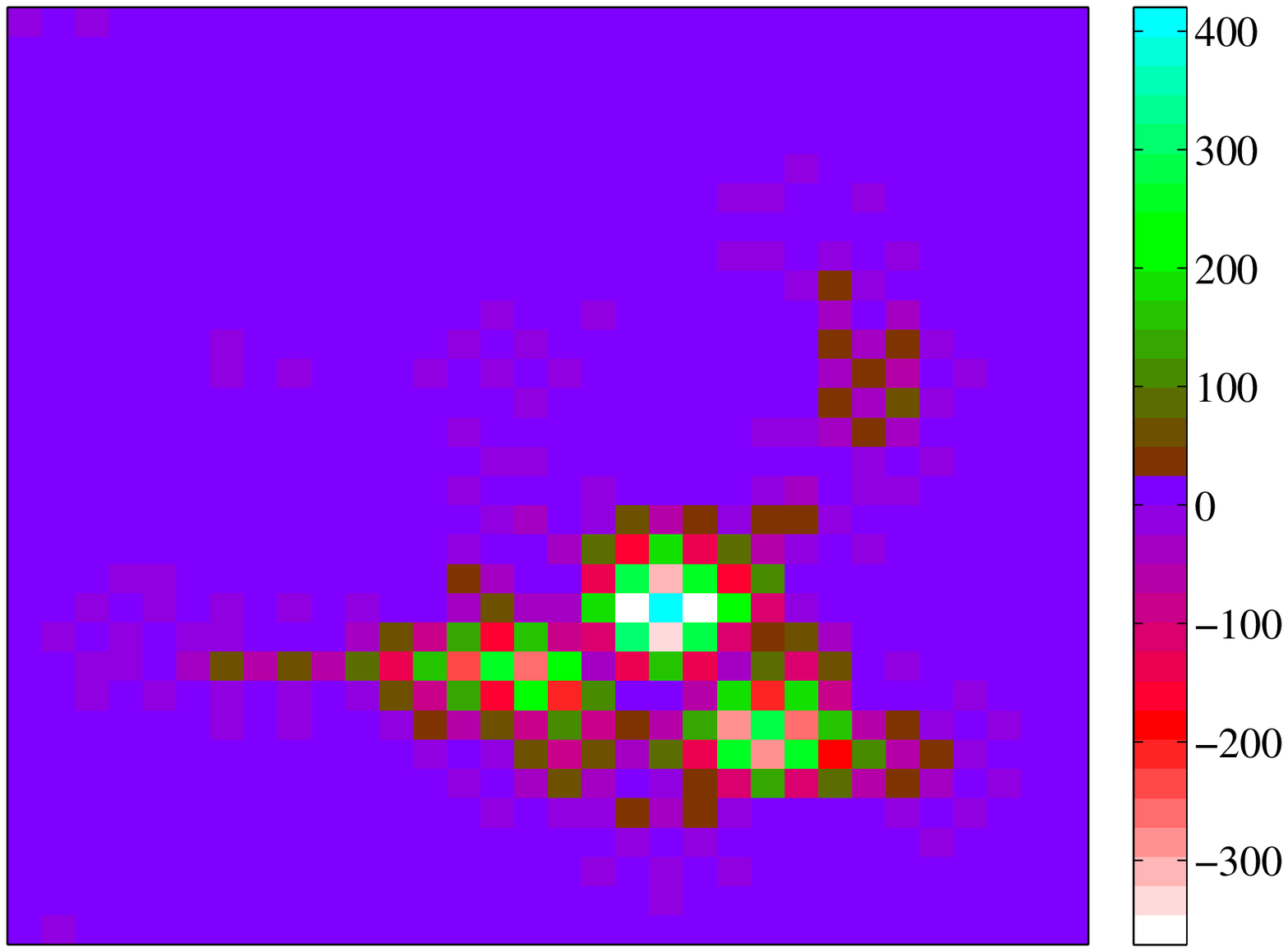,width=\FigFourSize}}
	\centerline{(b) STOMP (CFAR)}
\end{minipage}

\vspace{0.25cm}

%
%
\begin{minipage}[b]{0.48\linewidth}
	\centering
	\centerline{\epsfig{figure=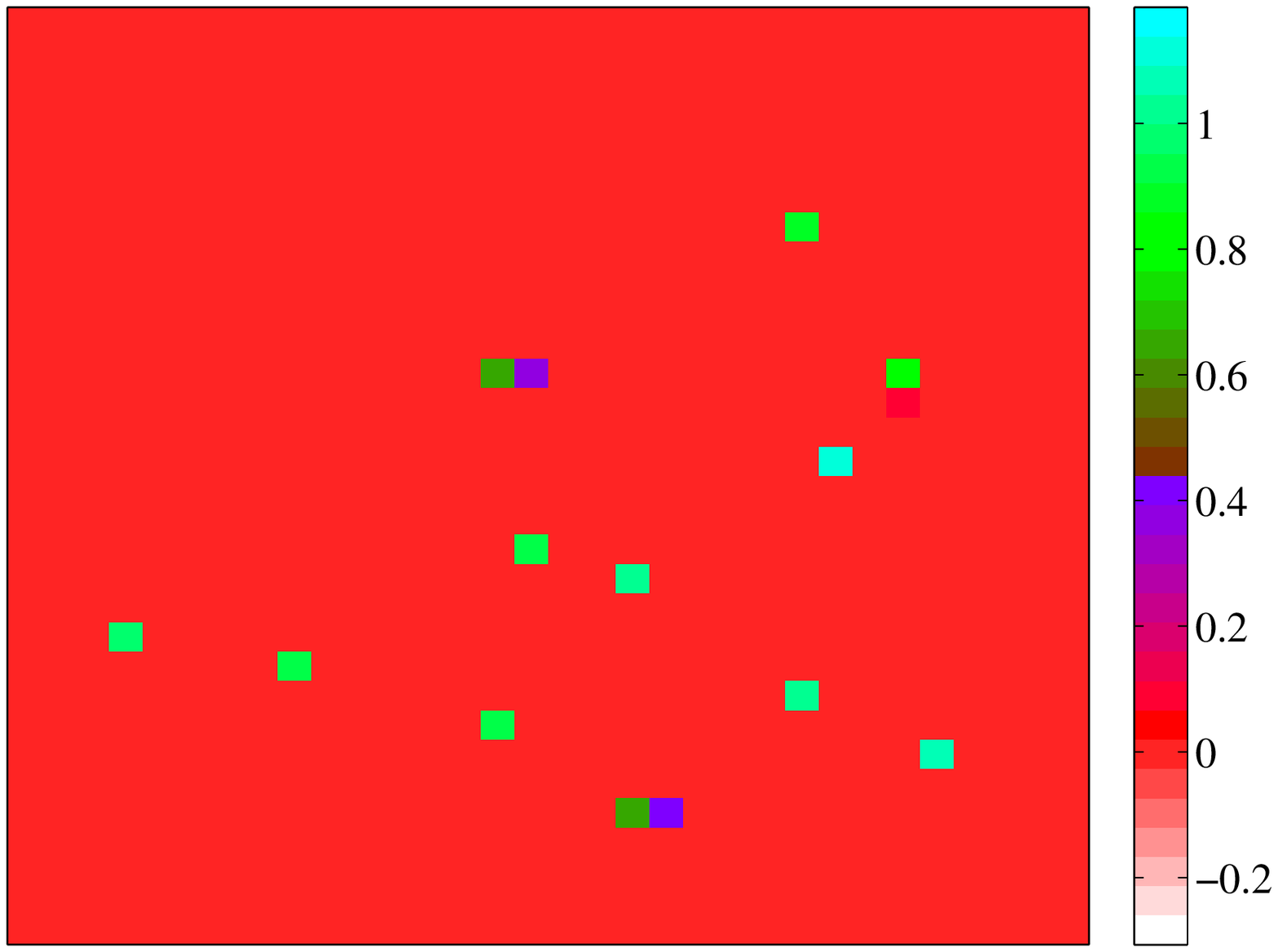,width=\FigFourSize}}
	\centerline{(c) MAP2 ($g^\ast = 1/\sqrt{2}$)}
\end{minipage}
\hfill
%
%
\begin{minipage}[b]{0.48\linewidth}
	\centering
	\centerline{\epsfig{figure=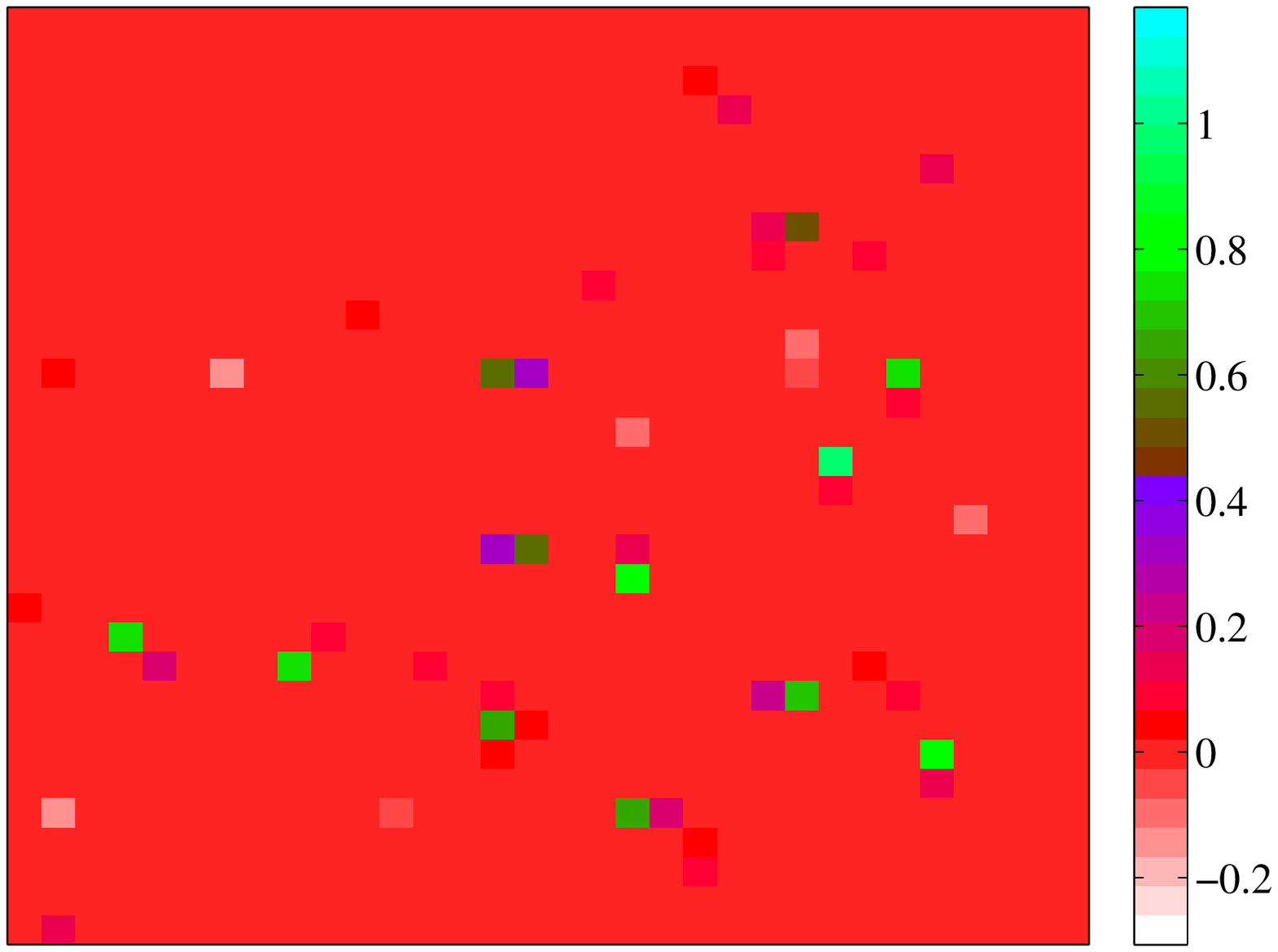,width=\FigFourSize}}
	\centerline{(d) lasso-SURE}
\end{minipage}

\caption{Reconstructed images for the binary-valued \underline{$\theta$}~under an SNR of 1.76 dB for SBL, StOMP (CFAR), MAP2 ($g^\ast=1/\sqrt{2}$), and lasso-SURE.}
\label{fig:binary:recon}
\end{figure}

In the low SNR case, MAP2 has the best performance. MAP1 consistently produces the \emph{trivial} estimate of all zeros, as evidenced by the mean value of $\|\hat{\vtor\theta}\|_0$ being equal to $0$. The trivial all-zero estimate results in $\|\vtor e\|_p=\|\vtor\theta\|_p$ for $p=0,1,2$. For a sparse $\vtor\theta$, a small $\|\vtor e\|_0$ therefore is not necessarily an indicator of good performance. 
A second comment regarding $\|\hat{\vtor\theta}\|_0$ is that it does not always give an accurate assessment of the \emph{perceived} sparsity of the reconstruction. In Table~\ref{tab:sim:binary}, SBL never produces a strictly sparse estimate, as the mean $\|\hat{\vtor\theta}\|_0$ equals the maximal value of 1024. However, consider Fig.~\ref{fig:binary:recon}a, where the SBL estimate for one noise realization at an SNR of 1.76 dB is depicted. The $\hat{\vtor\theta}$ looks sparser than would be suggested by $\|\hat{\vtor\theta}\|_0=1024$. This is because many of the non-zero pixel values have a small magnitude, and are visually indistinguishable from zero. 
The SBL estimate has many spurious non-zero pixels, in addition to blurring around several non-zero pixel locations. Negative values are present in the reconstruction, although the binary $\vtor\theta$ is non-negative.
%
%


The StOMP$^\text{CFAR}$, MAP2, and lasso-SURE estimate are illustrated in Figs.~\ref{fig:binary:recon}b--d respectively. The StOMP$^\text{CFAR}$ $\hat{\vtor\theta}$ has large positive and negative values. It does not seem like a sufficient number of stages have been taken. 
While blurring around several non-zero voxels are evident in the MAP2 estimate, $\vtor{\hat\theta}$ closely resembles $\vtor\theta$, cf.~Fig.~\ref{fig:numerical_study}a. None of the estimators considered here take into account positivity. From Fig.~\ref{fig:binary:recon}b, however, one sees that the MAP2 estimate has no negative values.
Qualitatively, the lasso-SURE estimate looks better than SBL, but worse than MAP2. This is reflected in the quantitative performance criteria in Table~\ref{tab:sim:binary}.


In the high SNR case, H-SURE has the best performance. The mean values of all the performance criteria decrease as compared to lasso-SURE. The greatest decreases are in $\|\vtor e\|_0$, $E_d$, and $\|\hat{\vtor\theta}\|_0$. They indicate that the H-SURE estimator is properly zeroing out spurious non-zero values and producing a sparser estimate than lasso-SURE. However, this comes at a price of higher computational complexity.


Examine next the performance of the reconstruction methods with the LAZE image. One expects MAP1 and MAP2 to have better performance than the other methods, as the image $\vtor\theta$ is generated using the LAZE prior. 
The numbers for the performance criteria are given in Table~\ref{tab:sim:laz}. Again, the reconstruction method with the best number for each criterion is underlined. For the LAZE $\vtor\theta$, $\|\vtor\theta\|_0=27$. 
%
%
\input{sim_img1_tip_tab2}



In the low SNR case, MAP2 has the advantage. MAP1 produces the trivial estimate of all zeros, just as in the case of the binary-valued $\vtor\theta$. The high SNR case has mixed results. While SBL has the best mean $\|\vtor e\|_1$ and $\|\vtor e\|_2$, the best result for the other three criteria each occur at a different method.
%
The fact that MAP1 and MAP2 did not produce superior performance over the other methods in the case of the LAZE image is unintuitive. As the SNR increases, however, the hyperparameter estimates become biased~\cite{tingthesis}.
The other unintuitive result is that $E_d(\vtor\theta,\hat{\vtor\theta};\delta)$ for the oracular LS estimate is not zero. This arises because of the choice of $\delta$. Since $\delta=10^{-2}\|\vtor\theta\|_\infty$, the values of $\hat{\theta}_i$ that are smaller than $\delta$ in absolute value are thresholded to zero. This results in a non-zero $E_d$ in some cases.

%

%

%% file: sim_img1_tip_tab1.tex
\begin{table}[!h]
\begin{flushleft}
\caption{Performance of the reconstruction methods for the binary-valued \underline{$\theta$}.}
\label{tab:sim:binary}
\end{flushleft}
\centering
\renewcommand{\arraystretch}{1.3} 
\begin{tabular}{|l|lllll|}
\hline 
& \multicolumn{5}{|c|}{\bfseries Error criterion} \\
\hline
\bfseries Method & $\|\vtor e\|_0$ & $\|\vtor e\|_1$ & $\|\vtor e\|_2$ & $E_d(\vtor\theta,\hat{\vtor\theta})$ & $\|\hat{\vtor\theta}\|_0$ \\
\hline \hline
\multicolumn{6}{|c|}{\bfseries SNR = 1.76 dB} \\
\hline
Oracular LS &
	12 & 0.880 & 0.309 & 0 & 12 \\
\hline
LS &
	1024 & 579 & 22.6 & $1.00 \! \times \! 10^3$ & 1024 \\
SBL &  
	1024 & 13.8 & 2.35 & 58.7 & 1024 \\
StOMP$^\text{CFAR}$ &
	335 & $1.46 \! \times \! 10^4$ & $1.50 \! \times \! 10^3$ & 322 & 335 \\
StOMP$^\text{CFDR}$ &
	454 & $7.95 \! \times \! 10^4$ & $7.17 \! \times \! 10^3$ & 442 & 454 \\
MAP1 &
	\best{12} & 12 & 3.46 & 12 & 0 \\
MAP2 &
	15.5 & \best{2.72} & \best{0.912} & \best{3.68} & \best{15.3} \\
lasso-SURE &
	60 & 7.83 & 1.51 & 44.2 & 60.6 \\
H-SURE &
	39.3 & 7.25 & 1.51 & 27.0 & 39.3 \\
\hline
\multicolumn{6}{|c|}{\bfseries SNR = 20 dB} \\
\hline
Oracular LS &
	12 & 0.112 & 0.0394 & 0 & 12 \\
\hline
LS &
	1024 & 86.1 & 3.67 & 929 & 1024 \\
SBL &
	1024 & 1.19 & 0.184 & 32.2 & 1024 \\
StOMP$^\text{CFAR}$ &
	377 & $4.33 \! \times \! 10^3$ & 457 & 361 & 377 \\
StOMP$^\text{CFDR}$ &
	459 & $1.36 \! \times \! 10^4$ & $1.19 \! \times \! 10^3$ & 446 & 459 \\
MAP1 &
	43.9 & 1.07 & 0.209 & 22.9 & 43.9 \\
MAP2 &
	230 & 3.82 & 0.380 & 114 & 230 \\
lasso-SURE &
	61.2 & 0.923 & 0.176 & 15.7 & 61.8 \\
H-SURE &
	\best{22.0} & \best{0.584} & \best{0.152} & \best{7.5} & \best{22.0} \\ 
\hline
\end{tabular}
\end{table}

%% file: sim_img1_tip_tab2.tex
\begin{table}[!h]
\begin{flushleft}
\caption{Performance of the reconstruction methods for the LAZE \underline{$\theta$}.} 
\label{tab:sim:laz}
\end{flushleft}
\centering
\renewcommand{\arraystretch}{1.3}
\begin{tabular}{|l|lllll|}
\hline 
& \multicolumn{5}{|c|}{\bfseries Error criterion} \\
\hline 
\bfseries Method & $\|\vtor e\|_0$ & $\|\vtor e\|_1$ & $\|\vtor e\|_2$ & $E_d(\vtor \theta,\hat{\vtor\theta})$ & $\|\hat{\vtor\theta}\|_0$ \\
\hline \hline
\multicolumn{6}{|c|}{\bfseries SNR = 1.76 dB} \\
\hline
Oracular LS &
	27 & 5.71 & 1.55 & 0.56 & 27 \\
\hline
LS &
	1024 & 807 & 31.6 & 977 & 1024 \\
SBL &  
	1024 & 28.1 & 3.99 & 72.6 & 1024 \\
StOMP$^\text{CFAR}$ &
	264 & $4.37 \! \times \! 10^3$ & 558 & 244 & 257 \\
StOMP$^\text{CFDR}$ &
	409 & $1.62 \! \times \! 10^4$ & $1.65 \! \times \! 10^3$ & 386 & 405 \\
MAP1 &
	\best{27} & 21.2 & 5.21 & 27 & 0 \\
MAP2 &
	30.9 & \best{17.5} & 3.98 & \best{25.1} & \best{9.77} \\
LASSO-SURE &
	92.6 & 20.3 & 3.15 & 69.3 & 81.9 \\
H-SURE &
	67.2 & 19.1 & \best{3.14} & 51.1 & 54.7 \\
\hline
\multicolumn{6}{|c|}{\bfseries SNR = 20 dB} \\
\hline
Oracular LS &
	27 & 0.686 & 0.190 & 0.6 & 27 \\
\hline
LS &
	1024 & 122 & 5.34 & 856 & 1024 \\
SBL &
	1024 & \best{4.32} & \best{0.814} & 33.7 & 1024 \\
StOMP$^\text{CFAR}$ &
	336 & $1.00 \! \times \! 10^4$ & 110 & 305 & 330 \\
StOMP$^\text{CFDR}$ &
	438 & $2.67 \! \times \! 10^4$ & 250 & 408 & 435 \\
MAP1 &
	\best{69.7} & 6.53 & 1.34 & 31.9 & \best{63.8} \\
MAP2 &
	216 & 10.8 & 1.44 & 86.6 & 212 \\
LASSO-SURE &
	119 & 6.63 & 1.32 & \best{31.1} & 116 \\
H-SURE &
	84.4 & 6.73 & 1.35 & 33.0 & 78.7 \\
\hline
\end{tabular}
\end{table}

%% file: sim_img2_tip.tex

\begin{figure}[!h]
\centering
\begin{minipage}[b]{0.48\linewidth}
	\centering
	\centerline{\epsfig{figure=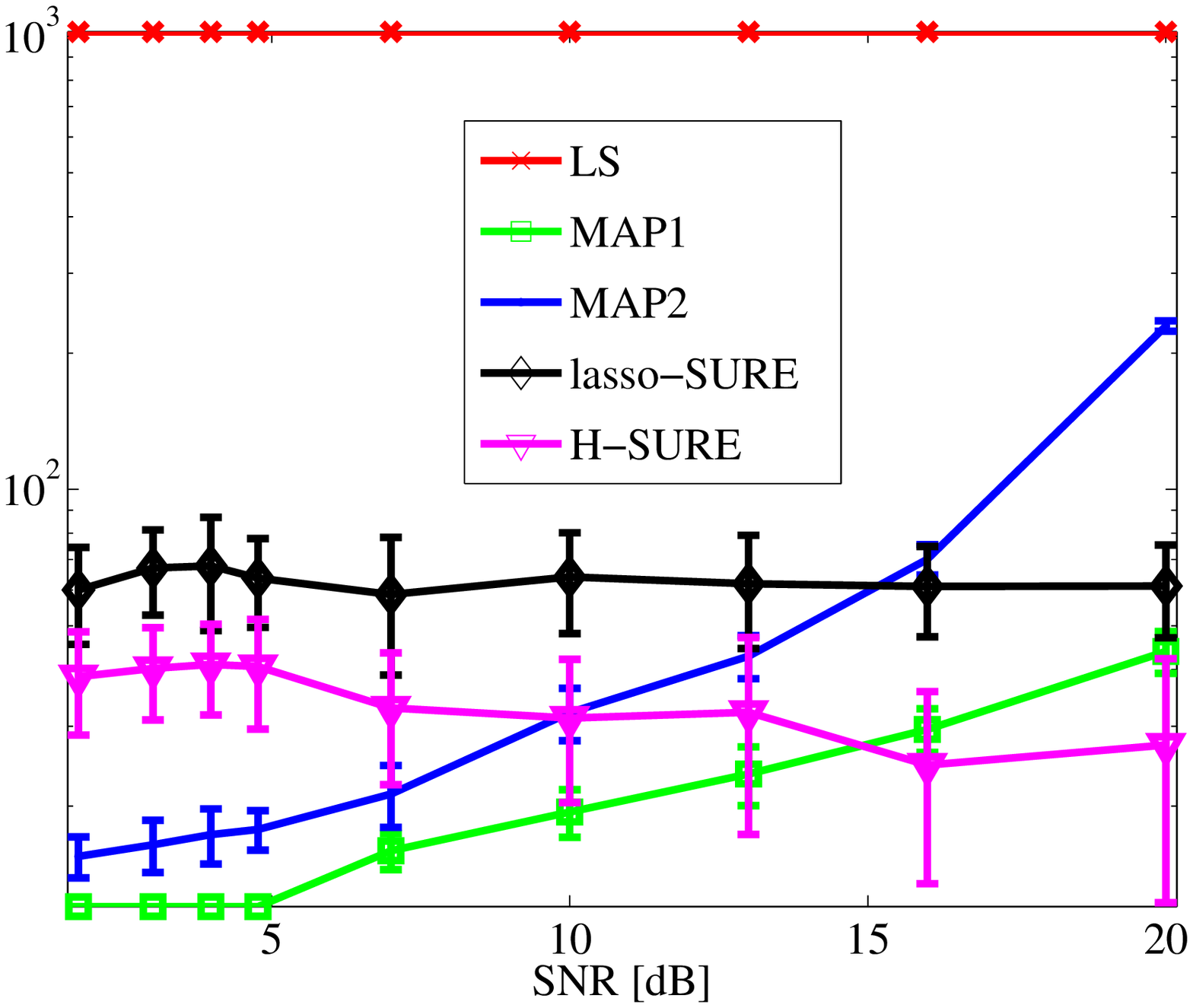,width=\FigSixSize}}
	\centerline{(a) $\|\vtor e\|_0$}
\end{minipage}
\hfill
\begin{minipage}[b]{0.48\linewidth}
	\centering
	\centerline{\epsfig{figure=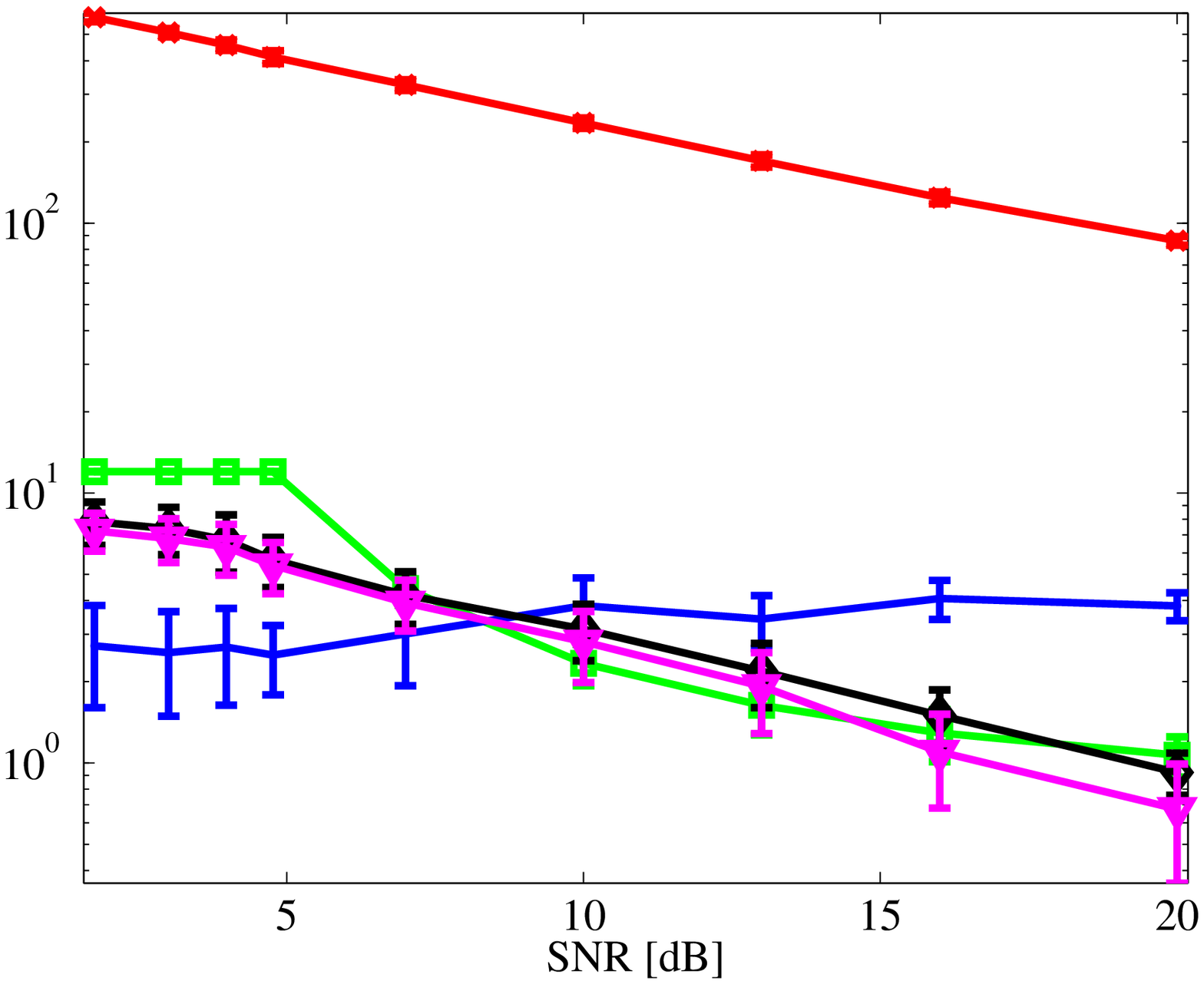,width=\FigSixSize}}
	\centerline{(b) $\|\vtor e\|_1$}
\end{minipage} 

\vspace{1em}

\begin{minipage}[b]{0.48\linewidth}
	\centering
	\centerline{\epsfig{figure=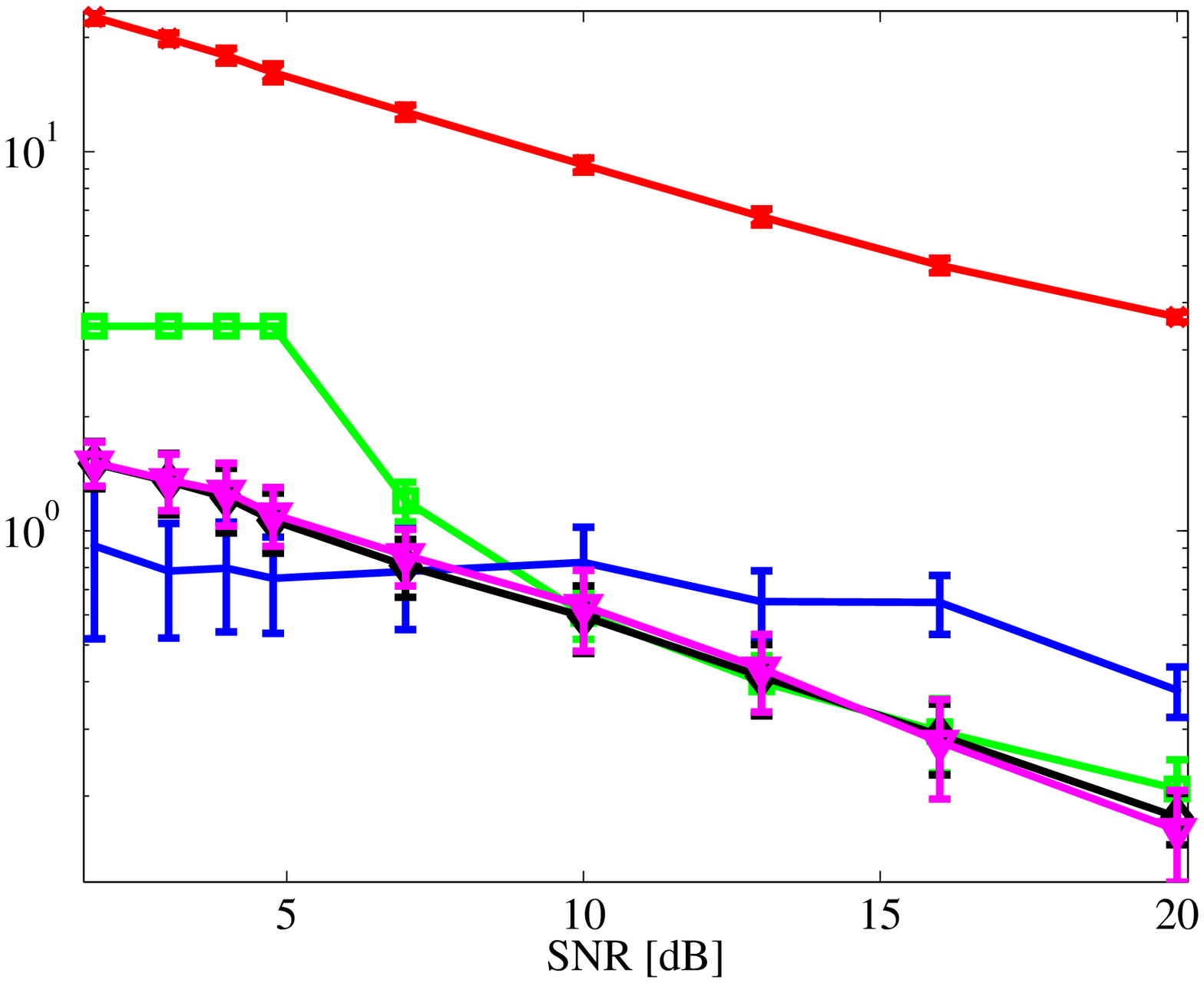,width=\FigSixSize}}
	\centerline{(c) $\|\vtor e\|_2$}
\end{minipage}
\hfill
\begin{minipage}[b]{0.48\linewidth}
	\centering
	\centerline{\epsfig{figure=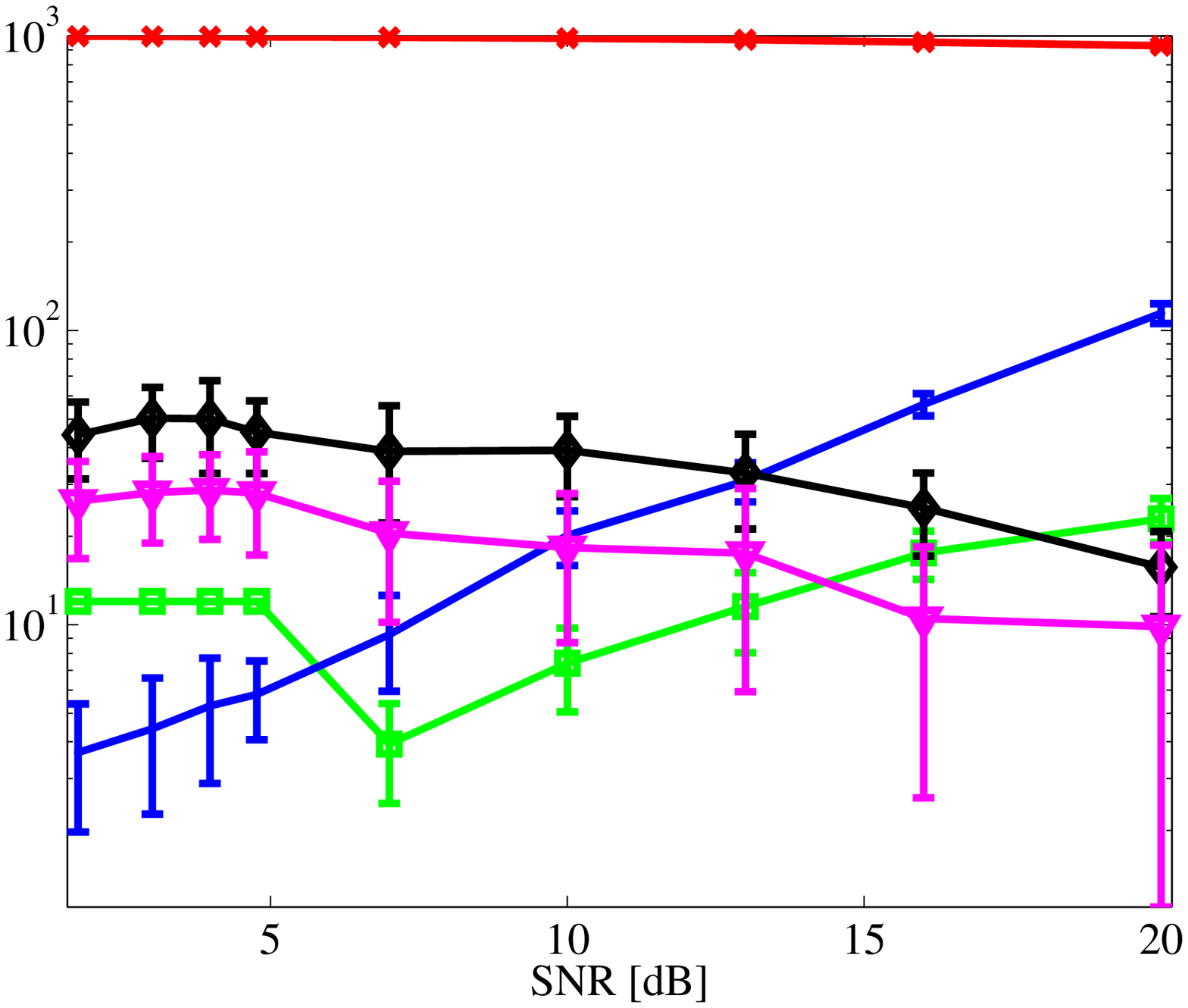,width=\FigSixSize}}
	\centerline{(d) $E_d$}
\end{minipage}

\vspace{1em}

\begin{minipage}[b]{0.48\linewidth}
	\centering
	\centerline{\epsfig{figure=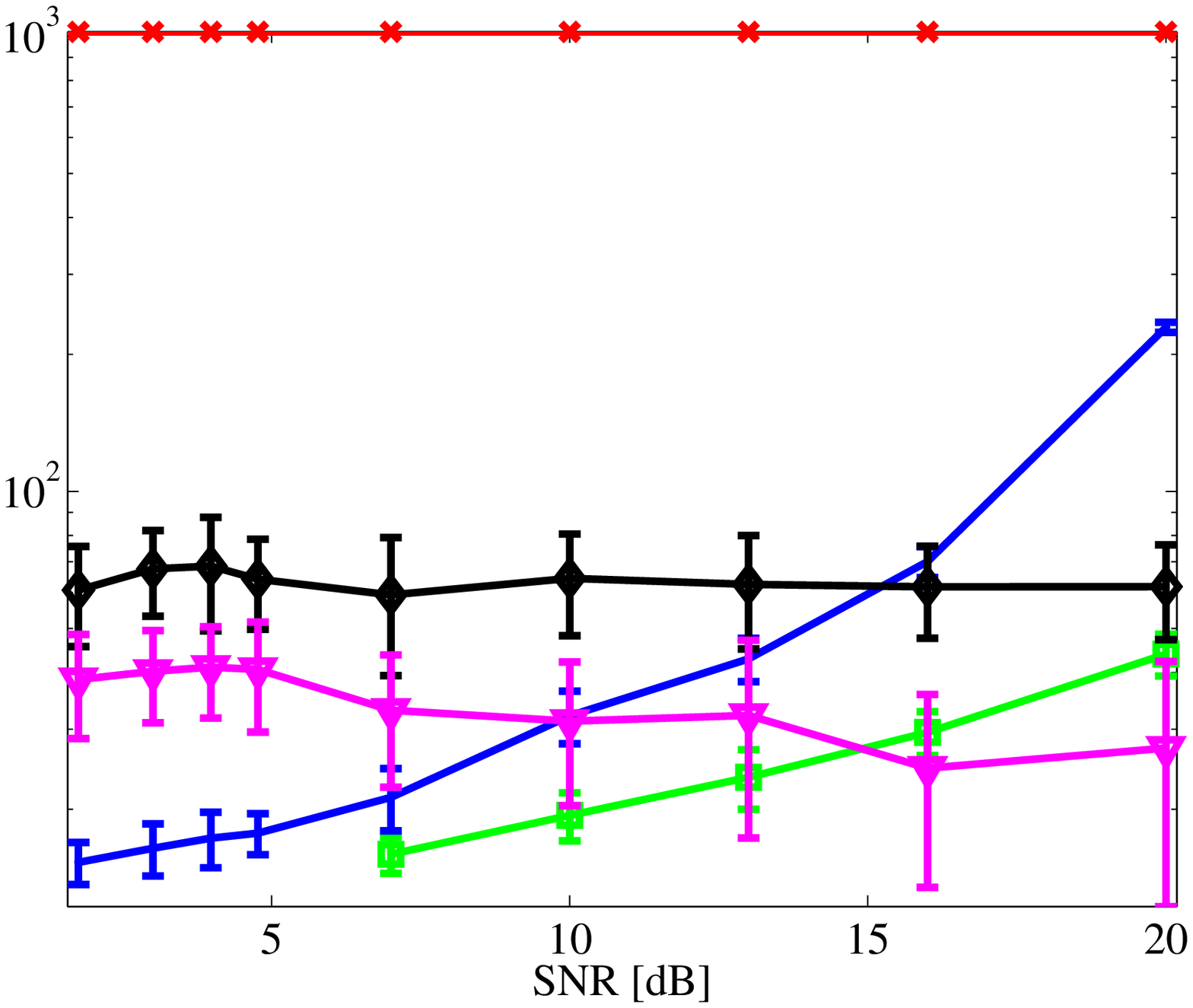,width=\FigSixSize}}
	\centerline{(e) $\|\hat{\vtor\theta}\|_0$}
\end{minipage}
\caption{Performance vs.~SNR for Landweber iterations, MAP1, MAP2, lasso-SURE, and H-SURE when applied to the binary-valued \underline{$\theta$}.}
\label{fig:snr_perf:1}
\end{figure}

The performance of the proposed reconstruction methods when applied to the binary-valued $\vtor\theta$ is examined with respect to SNR. The intent in this subsection is to study the behavior of the proposed methods at SNR values in between the low and high values of $1.76$ dB and $20$ dB respectively. As with the previous section, the MAP2 estimator is used with $g^\ast = 1/\sqrt{2}$. For each estimator, the mean is plotted along with error bars of one standard deviation. The error plots are given in Fig.~\ref{fig:snr_perf:1}.
%
%
Note that in Fig.~\ref{fig:snr_perf:1}e, the MAP1 curve is missing the first several SNR values because $\|\hat{\vtor\theta}\|_0=0$ and the y-axis is in a log scale.



First, consider the $\|\vtor e\|_0$, $\|\vtor e\|_1$, and $\|\vtor e\|_2$ error criteria. 
MAP1 is unable to distinguish the location of the non-zero pixels in low SNR. Under high SNR conditions, it has performance that is comparable to lasso-SURE and H-SURE in terms of the $\|\vtor e\|_1$ and $\|\vtor e\|_2$ errors. The value of $\|\vtor e\|_0$ increases with respect to increasing SNR for MAP1. Taken together with the $\|\vtor e\|_1$ and $\|\vtor e\|_2$ curves, the trend is indicative of small non-zero coefficients appearing in $\hat{\vtor\theta}$ that are spurious. 
MAP2 also has the same behavior with respect to $\|\vtor e\|_0$; however, a performance gap under high SNR exists in its $\|\vtor e\|_1$ and $\|\vtor e\|_2$ curves as compared to MAP1, lasso-SURE, and H-SURE. 
The lasso-SURE and H-SURE estimates have curves that decrease as the SNR increases. H-SURE's error curve is lower than lasso-SURE's for $\|\vtor e\|_0$ and $\|\vtor e\|_1$, and it is almost identical for $\|\vtor e\|_2$.


Consider next the $E_d$ and $\|\hat{\vtor\theta}\|_0$ error criterion. 
The lasso-SURE curve for $\|\hat{\vtor\theta}\|_0$ is relatively flat, and its $E_d$ curve decreases for high SNR. This indicates that, while the number of non-zero coefficients in $\hat{\vtor\theta}$ remains the same, the amplitude at the spurious locations are decreasing. 
With MAP1 and MAP2, the opposite trend is true. For low SNR, the number of non-zero coefficients in $\hat{\vtor\theta}$ is small, but increases with higher SNR. A similar increase can be seen in the $E_d$ curves. One can conclude that the number of spurious non-zero locations is increasing. This phenomenon is due to the bias of the hyperparameter estimates~\cite{tingthesis}. 
With H-SURE, both the $E_d$ and $\|\hat{\vtor\theta}\|_0$ curves decrease as the SNR increases. This behavior is intuitive, as higher SNR should result in better performance. We note that H-SURE's $E_d$ curve is lower than lasso-SURE's; moreover, H-SURE's $\|\hat{\vtor\theta}\|_0$ curve is closer to $\|\vtor\theta\|_0=12$ than lasso-SURE's.

%% file: sim_img4_tip.tex

\begin{figure}[!h]
\begin{minipage}[b]{0.48\linewidth}
	\centering
	\centerline{\epsfig{figure=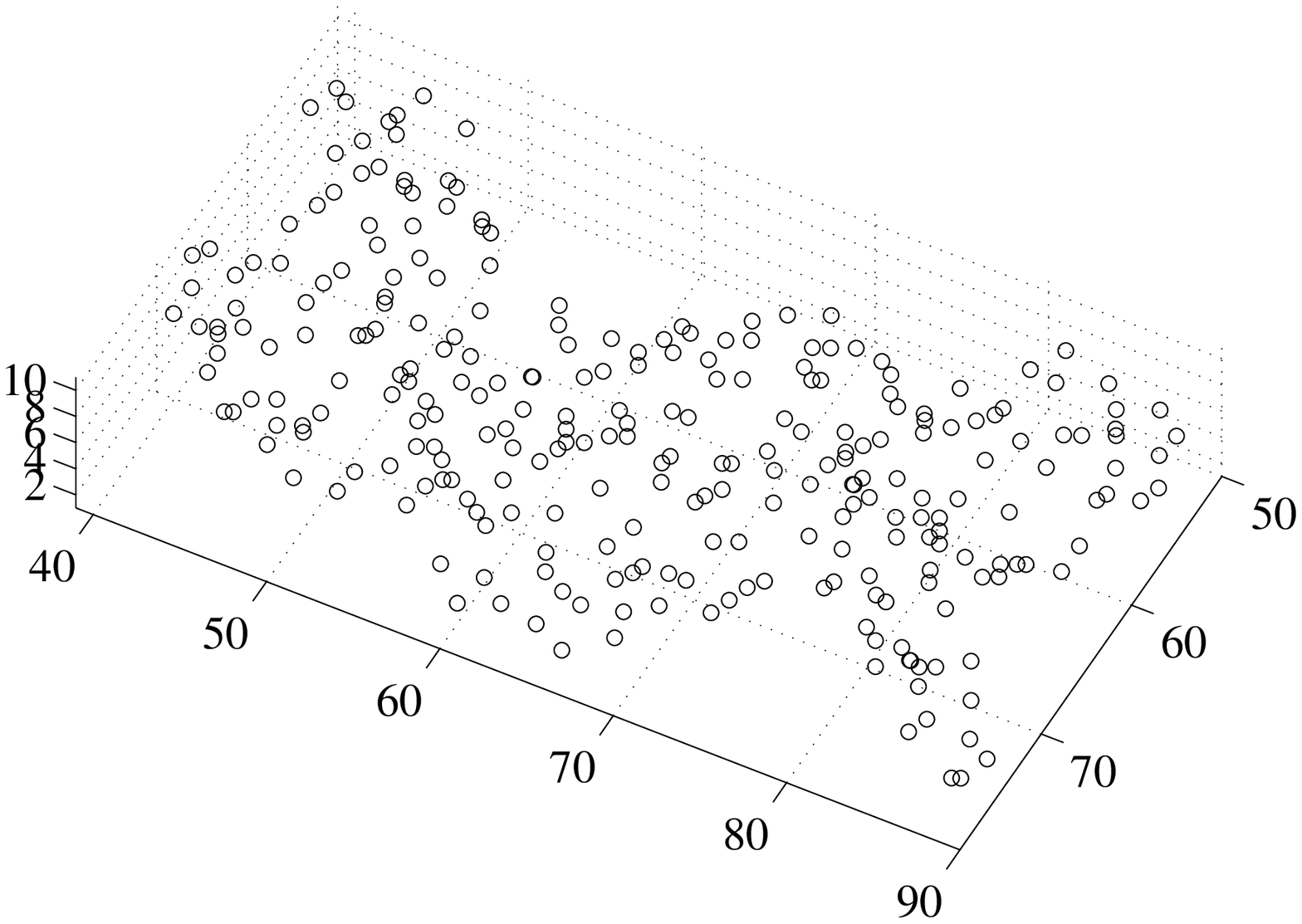,width=\FigNineSize}}
	\centerline{(a) Image $\vtor\theta$}
\end{minipage}
\hfill
\begin{minipage}[b]{0.48\linewidth}
	\centering
	\centerline{\epsfig{figure=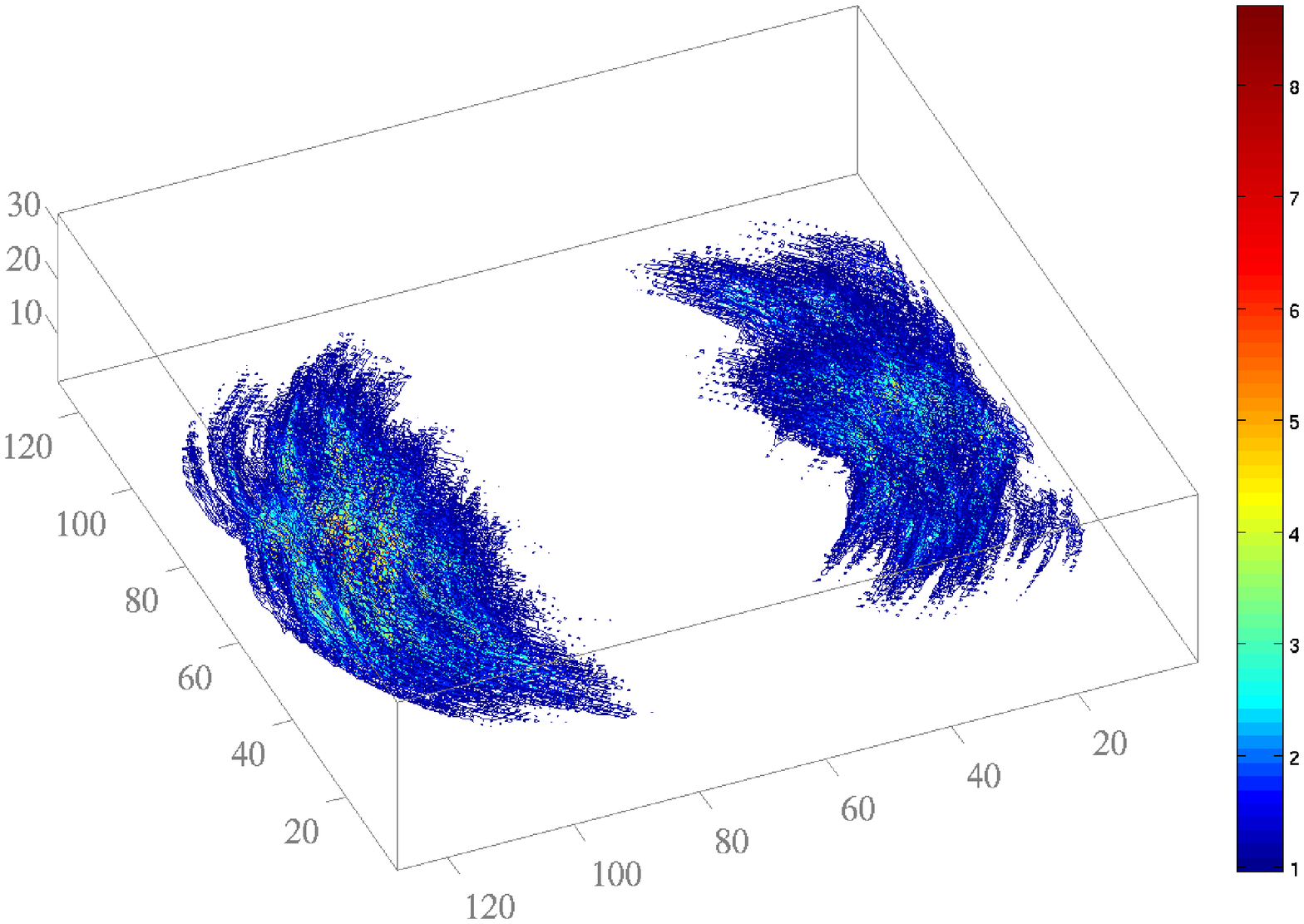,width=\FigNineSize}}
	\centerline{(b) Image $\mathbf{H}\vtor\theta$}
\end{minipage}
\caption{Image $\vtor\theta$ and noiseless projection $\mathbf{H}\vtor\theta$ used in the MRFM reconstruction example.}
\label{fig:mrfm:intro}
\end{figure}

\begin{figure}[!h]
	\centering
	\centerline{\epsfig{figure=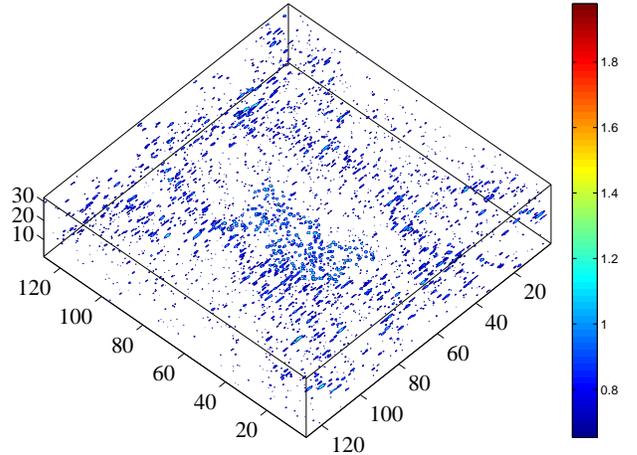,width=3.2in}}
	\caption{The LS estimate of the MRFM example under a SNR of 4.77 dB.}
	\label{fig:mrfm:recon_ls}
\end{figure}

\begin{figure}[!h]
	\centering
	\centerline{\epsfig{figure=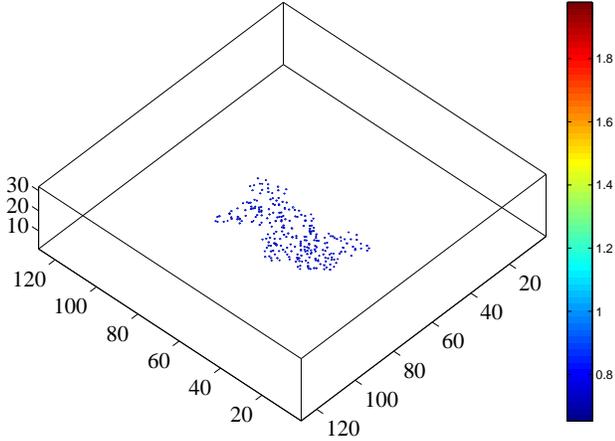,width=3.2in}}
	\caption{The lasso-SURE estimate of the MRFM example under a SNR of 4.77 dB.}
	\label{fig:mrfm:recon_lassos}
\end{figure}

\begin{figure}[!h]
\begin{minipage}[b]{0.48\linewidth}
	\centering
	\centerline{\epsfig{figure=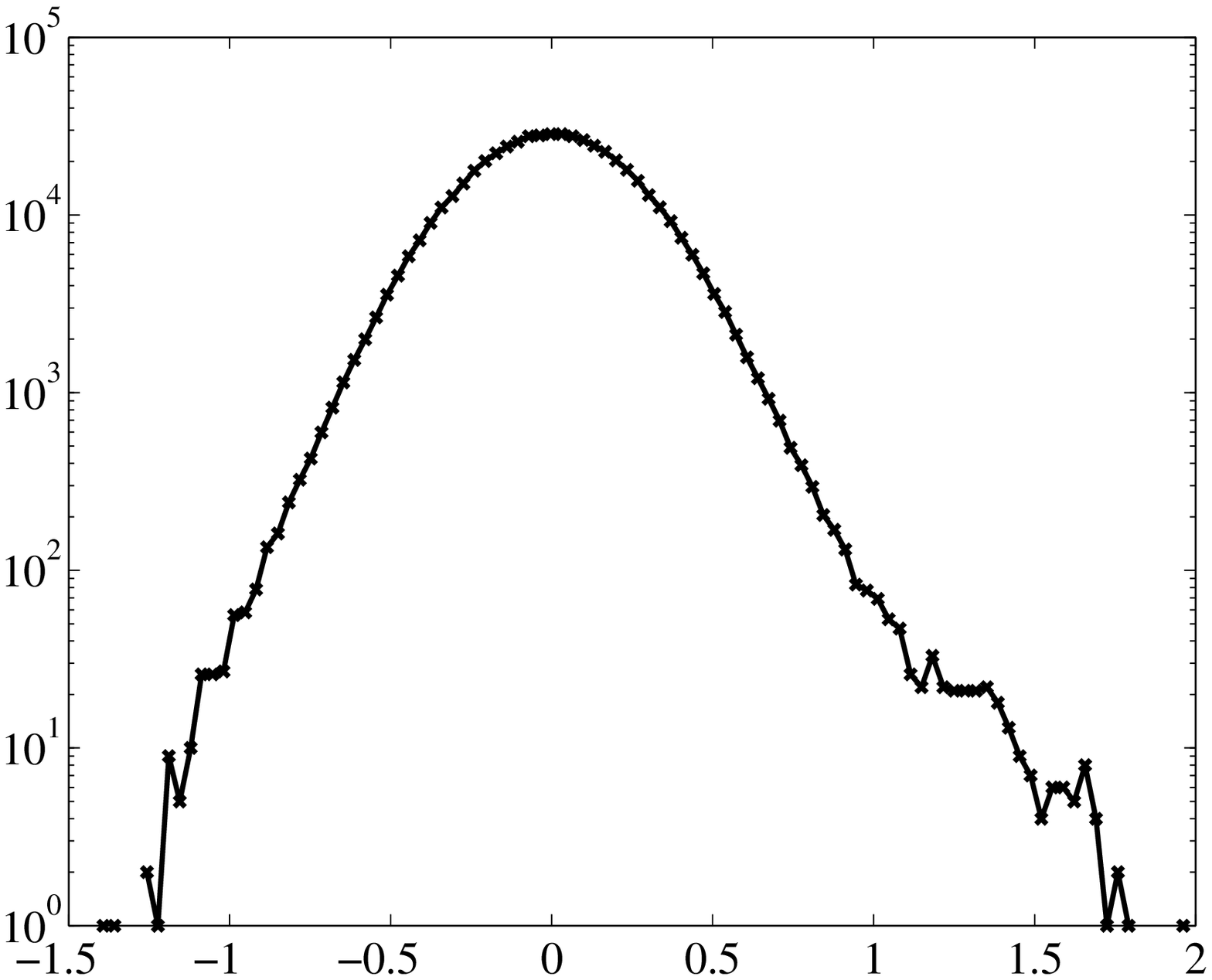,width=\FigTwelveSize}}
	\centerline{(a) LS}
\end{minipage}
\hfill
\begin{minipage}[b]{0.48\linewidth}
	\centering
	\centerline{\epsfig{figure=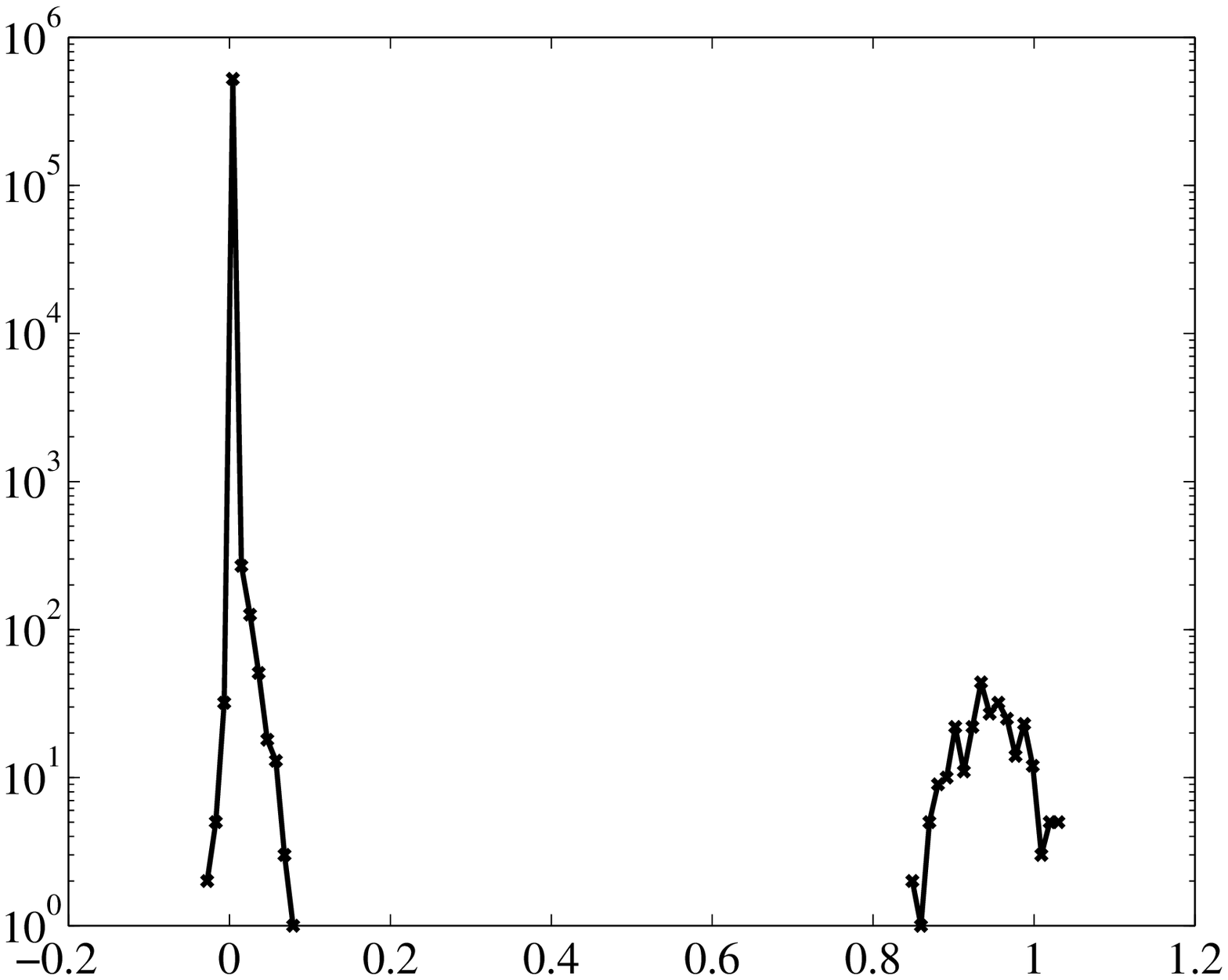,width=\FigTwelveSize}}
	\centerline{(b) lasso-SURE}
\end{minipage}
\caption{Histogram of $\hat{\theta}_i$ for the LS and lasso-SURE estimator.}
\label{fig:mrfm:hist}
\end{figure}

A three dimensional example using the hydrogen atom locations of the DNA molecule (PDB ID: 103D)~\cite{chou94newer} as $\vtor\theta$ and the 3d MRFM psf is carried out in this subsection. Both $\vtor \theta$ and $\vtor y$ have dimension $128 \times 128 \times 32$, and the SNR is 4.77 dB. Each hydrogen location in $\vtor\theta$ is set to 1, and the rest of the locations set to 0. The resulting image $\vtor\theta$ has a helical structure: see Fig.~\ref{fig:mrfm:intro}a. The image represented by $\mathbf{H}\vtor\theta$ is illustrated in Fig.~\ref{fig:mrfm:intro}b.
%
%
The LS and lasso-SURE estimates are given in Fig.~\ref{fig:mrfm:recon_ls} and~\ref{fig:mrfm:recon_lassos} respectively. 
%
%
%
%
The 3d figures plot contours for several values. The white volume in Fig.~\ref{fig:mrfm:recon_ls} does not indicate $\hat{\theta}_i=0$; rather, the $\hat{\theta}_i$ are at a value smaller than the lowest color bar value. On the other hand, the white volume of the lasso-SURE estimate is mostly $\hat{\theta}_i=0$. The histogram of $\hat{\theta}_i$ for the LS and lasso-SURE estimator given in Fig.~\ref{fig:mrfm:hist}a,b respectively illustrate this point.
%
%
The sharp peak at $0$ in the lasso-SURE histogram suggests that the lasso estimator incorporates a thresholding rule, which it does. The $\hat{\theta}_i$ values are separated into two distinct sets: the sparse image centered around $0.95$ and the background around 0. In contrast, the histogram of $\hat{\theta}_i$ for Landweber is not separated in this fashion, nor does it have a sharp peak at 0.

%% file: summary.tex

Use of a mixed discrete-continuous LAZE prior and jointly estimating $(\vtor\theta,\vtor\zeta)$ as the maximizer of $p(\vtor y,\vtor\theta|\vtor\zeta)$ gives rise to the Bernoulli-Laplacian sparse estimators MAP1 and MAP2. The hybrid thresholding rule is observed in both of these sparse estimators. When used in the iterative thresholding framework, the resulting penalty on $\vtor\theta$ is quadratic around the origin, and linear away from the origin, cf.~(\ref{eqn:hybrid:cost}). In order to apply lasso and the hybrid estimator to data, an empirical means of estimating the hyperparameters is required. This is achieved via Stein's unbiased risk estimate.

A numerical study shows that MAP1 and MAP2 perform well at low SNR, but the performance deteriorates at higher SNR. While StOMP demonstrates competitive results in~\cite{donoho06_stomp}, such is not the case in the simulation study conducted in this paper. The SBL estimate is not sparse; despite this, the estimates look visually sparse due to many non-zero values being small. In the high SNR regime for the LAZE $\vtor\theta$, SBL has good performance. When the hyperparameters are estimated via SURE, the hybrid estimator achieves a sparser estimate with lower $l_p$ reconstruction error for $p=0,1,2$ as compared to lasso. In addition, the hybrid estimator has lower detection error $E_d$. The numerical study suggests that sparse estimators based on sparse priors may achieve superior performance to the lasso.




The paper did not compare the MAP/ML and SURE estimates of the hyperparameters to other estimates, e.g., GCV, the method of~\cite{yuan05} for lasso, etc. This is primarily due to a lack of space. In the case when $\vtor{\hat\theta}$ is a linear function of $\vtor y$, SURE is equivalent to the $C_p$ statistic, while GCV is the $C_p$ statistic with $\sigma^2$ replaced by an estimated version~\cite{efron01}. Unfortunately, the sparse estimators considered in the paper are all nonlinear in $\vtor y$.
Another issue that should be looked in future work is how to improve MAP1/2 to rectify the deteriorating performance at higher SNR. The estimates $\hat{a},\hat{w}$ generally become more biased as the SNR increases~\cite{tingthesis}. This has been noted in~\cite{mackay99}. With MAP2, the degree of bias is affected by the selection of $g^\ast$. 


Implementation considerations were not discussed, although they are critical in the implementation of a deconvolution algorithm. The interested reader is referred to~\cite{tingthesis}. In terms of increasing complexity, the estimators can be approximately ordered as: StOMP, LS/oracular LS, MAP1 and MAP2, lasso-SURE, H-SURE, and SBL. 
Thanks to LARS, evaluating a goodness-of-fit criterion for lasso whether it be a SURE criterion, a GCV criterion, etc.~has low computational complexity. Although LARS requires the selection of individual columns of $\mathbf{H}$, this is not an issue when $\mathbf{H}$ represents a convolution operator. The selection can be efficiently implemented using the fast Fourier transform (FFT). 
Solving for the H-SURE hyperparameters has higher computational complexity since an efficient implementation of the H-SURE estimator is currently lacking. In this paper, the iterative thresholding framework is used for part of the solution; however, a LARS-like method would be a welcomed improvement.

%% file: proof_iv.tex
A more general result is derived here. Consider the iteration
%
%
\begin{equation} \label{eqn:em_iter:general}
 \vtor{\hat{\theta}}^{(n+1)} = S_{T,\vtor\zeta} \left( \vtor{\hat{\theta}}^{(n)} +
	(\alpha/\sigma)^2 \mathbf{H}^\prime ( \vtor y - \mathbf{H} \vtor{\hat{\theta}}^{(n)} \right),
\end{equation}
where $T(\vtor x;\vtor \zeta) = \sum_i T(x_i;\vtor\zeta) \vtor e_i$ is a thresholding rule~\cite[Sec.~2.3]{johnstone04} with the following condition. Suppose that $T(\cdot;\vtor\zeta)$ has \emph{threshold} $t>0$; then, $T(\cdot;\vtor\zeta)$ is strictly increasing on $\mathbb{R} \setminus (-t,t)$. Note that $T^{-1}(x;\vtor\zeta)$ is only defined for $x \neq 0$. Extend the definition at $x=0$ to get 
%
%
\begin{equation} \label{eqn:t_inverse}
 T^\dagger(x;\vtor\zeta) = \left\{ \begin{array}{cc}
				0 & x = 0 \\
				T^{-1}(x;\vtor\zeta) & x \neq 0
				\end{array} \right.
\end{equation}
$T^\dagger(x;\vtor\zeta)$ is continuous on $x \in \mathbb{R} \setminus \{0\}$. For the remainder of this section, the dependency of $T$, $T^{-1}$, and $T^\dagger$ on $\vtor\zeta$ will be omitted for the sake of brevity.

%
%
\begin{prop} \label{prop:j1}
The function
%
%
\begin{equation} \label{eqn:j1}
 J_1(x) \triangleq 2 T^\dagger(x)x - x^2 - 2 \int_0^\xi T(q) \ud q \bigg\vert_{\xi=T^\dagger(x)}
\end{equation}
is continuous for $x \in \mathbb{R}$.
\end{prop}

\textbf{Proof}. Since $T^\dagger(x)$ is continuous in $\mathbb{R} \setminus \{0\}$, the only place that should be checked is $x=0$. The second term in (\ref{eqn:j1}) is continuous, so it remains to check the first and third terms. By definition of a threshold function, $T^\dagger(0^+)=t$ and $T^\dagger(0^-)=-t$.

Consider $\epsilon>0$. Since $T^\dagger(\cdot)$ is right continuous at $0^+$, there exists $\delta_1>0$ s.t.~$x \in (0,\delta_1)$ implies that $|T^\dagger(x)-t|<0.1t$. Likewise, since $T^\dagger(\cdot)$ is left continuous at $0^-$, there exists $\delta_2>0$ s.t.~$x \in (-\delta_2,0)$ implies that $|T^\dagger(x)+t|<0.1t$. Set
%
%
\begin{equation*}
 \delta=\frac{1}{2}\min(\delta_1,\delta_2,\frac{\epsilon}{1.1t})
\end{equation*}
so that $|x|<\delta \Longrightarrow |xT^\dagger(x)|<\epsilon$. 

Consider the third term. Define $A(\xi) \triangleq \int_0^\xi T(q) dq$: since $T(\cdot)$ is continuous, so is $A(\cdot)$. Moreover, for $|x| \le t$, $A(x)=0$. For $\epsilon>0$, there exists $\kappa > t$ s.t.~$|\xi|<\kappa \Longrightarrow |A(\xi)| < \epsilon$. Since $T^\dagger(\cdot)$ is right continuous at $0^+$, there exists $\delta_1>0$ s.t.~$x \in (0,\delta_1) \Longrightarrow |T^\dagger(x)-t|<\kappa-t$. In a similar fashion, since $T^\dagger(\cdot)$ is left continuous at $0^-$, there exists $\delta_2>0$ s.t.~$x \in (-\delta_2,0) \Longrightarrow |T^\dagger(x)+t|<\kappa-t$. Set $\delta=\min(\delta_1,\delta_2)$. From $|x|<\delta$, one gets $|T^\dagger(x)|<\kappa$ whence $|A(T^\dagger(x))|<\epsilon$. $\blacksquare$

%
%
\begin{prop} \label{prop:j1_optim}
The minimizer of $\varphi(x) = x^2 - 2cx + J_1(x)$ is $\tilde{x}=T(c)$.
\end{prop}

\textbf{Proof}. Let $\varphi_1(x) \triangleq x^2-2cx$: $\varphi_1^\prime(x)=2(x-c)$, and is lower bounded. Similarly, consider $J_1(x)$: for $x \neq 0$, $J_1^\prime(x) = 2(T^\dagger(x)-x)$. Since $0 \le T(x) \le x$ for all $x \ge 0$ and $x \le T(x) \le 0$ for all $x < 0$,
%
%
\begin{equation*}
 J_1^\prime(x) = \left\{ \begin{array}{cc}
                          \ge 0 & x > 0 \\
			  \le 0 & x < 0 \\
                         \end{array} \right.
\end{equation*}
$J_1(x)$ is also lower bounded. Applying Prop.~\ref{prop:j1} results in $\varphi(x)$ being a continuous, lower bounded fuction. Consider now two cases.

Case 1: $|c|>t$, where recall that $t$ is the threshold of $T(\cdot)$. For $x \neq 0$, $\varphi^\prime(x)=2x-2c+J_1^\prime(x)=2[T^\dagger(x))-c]$. So $\varphi^\prime(x)=0$ iff $T^\dagger(x)=c$, which occurs uniquely at $\tilde{x}=T(c)>0$. Consider 
%
%
\begin{equation} \label{eqn:proof_iv:p2_1}
 \varphi^\prime(T(c)+\delta)=2[T^\dagger(T(c)+\delta)-c]
\end{equation}
Since we assume that $T(\cdot)$ is strictly increasing on $\mathbb{R} \setminus (-t,t)$, $T^\dagger(x)$ is also strictly increasing for $x \neq 0$. For sufficiently small $\delta>0$, $\varphi^\prime(T(c)+\delta)>0$ and $\varphi^\prime(T(c)-\delta)<0$. So $\tilde{x}=T(c)$ is a local minimum. At this value of $x$, $\varphi(x)=-2A(c)<0$. To verify that $\tilde{x}$ is the global minimum, it is necessary to compute $\varphi(0)=0$. So indeed, $x=T(c)$ minimizes $\varphi(x)$.

Case 2: $|c| \le t$. Suppose that the minimizer $x \neq 0$. Then, the analysis in Case 1 applies, resulting in $x = T(c)$. But since $|c| \le t$ by assumption, one gets $x=0$. This is a contradiction: it must therefore be the case that $\tilde{x}=0$. $\blacksquare$

%
%
\begin{theorem} \label{thm:general:cost}
Suppose that $\|\mathbf{H}\|_2 < 1$ and $\alpha=\sigma$. Consider the iteration (\ref{eqn:em_iter:general}), where $T(\cdot)$ is a thresholding rule with threshold $t>0$, and $T(\cdot)$ is strictly increasing in $\mathbb{R} \setminus (-t,t)$. Then, the iterations (\ref{eqn:em_iter:general}) converge to a stationary point of  $\Psi(\vtor\theta)$, where
%
%
\begin{align}
\Psi(\vtor\theta) & = \|\mathbf{H}\vtor\theta - \vtor y\|_2^2 + J(\vtor\theta) \nonumber \\
\text{ where: } J(\vtor\theta) & \triangleq \sum_{i=1}^M J_1(\theta_i) 
	\label{eqn:general:cost_fuc}
\end{align}
\end{theorem}

\textbf{Proof}. Use the following definitions, which appear in~\cite{daubechies04}:
%
%
\begin{align}
\Xi(\vtor\theta;\vtor a) & \triangleq C\|\vtor\theta-\vtor a\|_2^2 - 
	\|\mathbf{H}\vtor\theta-\mathbf{H}\vtor a\|_2^2
	\label{eqn:app_ebd:1} \\
\Phi^\text{SUR}(\vtor\theta;\vtor a) & \triangleq \Phi( \vtor\theta ) + \Xi(\vtor\theta;\vtor a),
	\label{eqn:app_ebd:2}
\end{align}
where $C$ is chosen to ensure that $\Xi(\vtor\theta;\vtor a)$ is strictly positive and convex in $\vtor\theta$ for any choice of $\vtor a$. By assumption, $\|H\|_2<1$, and so select $C=1$~\cite{daubechies04}. The function $\Phi^\text{SUR}(\vtor\theta;\vtor a)$ is the surrogate function that is minimized in place of $\Phi(\vtor\theta)$. Consider the minimization of $\Phi^\text{SUR}(\vtor\theta;\vtor a)$, which can be simplified as
%
%
\begin{multline}
\Phi^\text{SUR}(\vtor\theta;\vtor a) = \|\vtor\theta\|^2 - 2(\vtor a+\mathbf{H}^\prime(\vtor y-\mathbf{H}\vtor a))^\prime\vtor\theta + \\
	J(\vtor\theta) + \|\vtor y\|^2 + \|\vtor a\|^2 - \|\mathbf{H}\vtor a\|^2
	\label{eqn:proof_iv:1}
\end{multline}
Since $J(\vtor\theta) = \sum_i J_1(\theta_i)$, the minimization of $\Phi^\text{SUR}(\vtor\theta;\vtor a)$ can be decomposed into $M$ subproblems, where each $\theta_i$ is separately minimized. Indeed, each $\theta_i$ should minimize
%
%
\begin{equation} \label{eqn:proof_iv:2}
\varphi(\theta_i) \triangleq \theta_i^2 - 2 s_i \theta_i + J_1(\theta_i),
\end{equation}
where $\vtor s \triangleq \vtor a+\mathbf{H}^\prime(\vtor y-\mathbf{H}\vtor a)$. Apply Prop.~\ref{prop:j1_optim} to get the minimizing $\theta_i$, i.e., $\theta_i=T(s_i)$. 

Let $\vtor{\hat{\theta}}^{(n)}$ denote the sequence generated by 
%
%
\begin{equation} \label{eqn:surrogate}
 \vtor{\hat{\theta}}^{(n+1)} = \mathop{\text{argmin}}_{\vtor\theta}
	\Phi^\text{SUR}( \vtor\theta;\vtor{\hat{\theta}}^{(n)} )
\end{equation}
where $\vtor{\hat{\theta}}^{(0)}$ is the initial estimate. Then, $\vtor{\hat{\theta}}^{(n)}$ is generated by (\ref{eqn:em_iter:general}), where recall that $\alpha/\sigma=1$. Any limit point of the iterations (\ref{eqn:em_iter:general}) is a stationary point of (\ref{eqn:general:cost_fuc})~\cite{lange00}. $\blacksquare$

%% file: proof_v.tex
\subsection{Proof of Thm.~\ref{thm:sure1}}

Recall that $\mathbf{G}(\mathbf{H}) = \mathbf{H}^\prime\mathbf{H}$ is the Gram matrix of $\mathbf{H}$.
In order to simplify notation, for $\mathbf{A} \in \mathbb{R}^{M \times M}$, denote by $\mathbf{A}_{11}=\mathbf{A}[1:r,1:r]$, $\mathbf{A}_{12}=\mathbf{A}[1:r,r+1:M]$, $\mathbf{A}_{21}=\mathbf{A}[r+1:M,1:r]$, and $\mathbf{A}_{22}=\mathbf{A}[r+1:M,r+1:M]$. 
The following proposition is needed. Its proof is omitted due to a lack of space. 
\begin{prop} \label{prop:v_1}
If $\mathbf{H}$ has linearly independent columns,
%
%
\begin{equation} \label{eqn:prop_v1:1}
\text{det}((\mathbf{P}\mathbf{G}(\mathbf{H})\mathbf{P}^\prime)_{22}) \neq 0.
\end{equation}
where $\mathbf{P}$ is a matrix that orders the zero and non-zero components of $\vtor{\hat\theta}$.
\end{prop}
For $\vtor{\hat\theta}$, an unbiased estimate of the $l_2$ risk (\ref{eqn:risk_Htheta:sure}) is~\cite{stein81,solo96}
%
%
\begin{equation} \label{eqn:sure1:rhat_primitive}
\hat{R}(\vtor\zeta) = \sigma^2 + \frac{\|\vtor e\|_2^2}{N} - \frac{2\sigma^2}{N} \sum_{n=1}^N \frac{\partial e_n}{\partial y_n}
\end{equation}
where $\vtor e = \vtor y - \mathbf{H}\vtor{\hat\theta}$. If $\vtor{\hat\theta}$ is obtained via a minimization $\vtor{\hat\theta}=\text{argmin}_{\vtor\theta} \Psi_{\vtor\zeta}(\vtor\theta)$, (\ref{eqn:sure1:rhat_primitive}) can be evaluated as~\cite[(2)]{solo96}
%
%
\begin{equation} \label{eqn:sure1:rhat_inverse_prob}
 \hat{R}(\vtor\zeta) = \sigma^2 + \frac{\|\vtor e\|_2^2}{N} -
	\frac{2\sigma^2}{N} \text{tr}( \mathbf{H} 
	(\mathbf{D}_{\vtor\theta\vtor\theta}\Psi_{\vtor\zeta})^{-1} 
	\mathbf{D}_{\vtor\theta\vtor y}\Psi_{\vtor\zeta}) \bigg|_{\vtor\theta=\vtor{\hat\theta}},
\end{equation}
where $\mathbf{D}_{\vtor u,\vtor v}(\cdot) \triangleq \partial^2(\cdot)/\partial \vtor u \partial \vtor v^\prime$. 

Let $\Psi_{\vtor\zeta,l}(\vtor\theta) = \|\mathbf{H}\vtor\theta-\vtor y\|_2^2 + \zeta_1\|\vtor\theta\|_1$ denote the cost function of lasso. 
Since $\Psi_{\vtor\zeta,l}(\vtor\theta)$ is not twice differentiable on $\mathbb{R}^M$, (\ref{eqn:sure1:rhat_inverse_prob}) cannot be directly applied. Consider 
%
%
\begin{multline} \label{eqn:sure1:psi_relax}
 \Psi_{\vtor\zeta,l}(\vtor\theta;a) = \|\mathbf{H}\vtor\theta- \vtor y\|_2^2\\
+ \frac{2\zeta_1}{\pi} \sum_{m=1}^M \{ \theta_i \text{arctan}\big(\frac{\theta_i}{a}\big) - \frac{a}{2} \ln\big(1+\frac{\theta_i^2}{a^2}\big) \}
\end{multline}
which is twice differentiable on $\mathbb{R}^M$. It can be shown that $\lim_{a\rightarrow 0} \Psi_{\vtor\zeta,l}(\vtor\theta;a)=\Psi_{\vtor\zeta,l}(\vtor\theta)$ pointwise. The minimizer of $\Psi_{\vtor\zeta,l}(\vtor\theta;a)$ therefore equals the minimizer of $\Psi_{\vtor\zeta,l}(\vtor\theta)$ in the limit as $a \rightarrow 0$. Denote by $\hat{R}_l(\vtor\zeta;a)$ the unbiased estimate of (\ref{eqn:risk_Htheta:sure}) when $\hat{\vtor\theta}$ is obtained by minimizing $\Psi_{\vtor\zeta,l}(\vtor\theta;a)$. As the RHS of (\ref{eqn:sure1:rhat_primitive}) is solely a function of $\hat{\vtor\theta}$ (recall that $\vtor y$, $\mathbf{H}$, and $\sigma^2$ are known), $\lim_{a\rightarrow 0} \hat{R}_l(\vtor\zeta;a)=\hat{R}_l(\vtor\zeta)$ pointwise.

Applying (\ref{eqn:sure1:rhat_inverse_prob}) ,
%
%
\begin{equation} \label{eqn:sure1:rhat_relax}
 \hat{R}_l(\vtor\zeta;a) = \sigma^2 + \frac{\|\vtor e\|_2^2}{N} + 
	\frac{2\sigma^2}{N} \text{tr}(\mathbf{G}(\mathbf{H})[\mathbf{G}(\mathbf{H})+\frac{1}{2}\mathbf{Z}_a(\vtor{\hat\theta})]^{-1})
\end{equation}
where
%
%
\begin{equation} \label{eqn:sure1:za}
 \mathbf{Z}_a(\vtor{\theta}) \triangleq \frac{2\zeta_1}{\pi} \text{diag}(
	\frac{a}{a^2+\theta_1^2},\ldots,\frac{a}{a^2+\theta_M^2}).
\end{equation}
Consider the $\text{tr}(\cdot)$ expression in (\ref{eqn:sure1:rhat_relax}). As $\mathbf{P}$ is orthogonal and matrix multiplication is commutative under the trace operator, 
%
%
\begin{multline*}
 \text{tr}(\mathbf{G}(\mathbf{H})\left[\mathbf{G}(\mathbf{H}+
	\frac{\mathbf{Z}_a(\vtor{\hat\theta})}{2}\right]^{-1}) = \\
 \text{tr}(\mathbf{P}\mathbf{G}(\mathbf{H})\mathbf{P}^\prime\left[
	\mathbf{P}\mathbf{G}(\mathbf{H})\mathbf{P}^\prime+
	\frac{\mathbf{P}\mathbf{Z}_a(\vtor{\hat\theta})\mathbf{P}^\prime}{2}
	\right]^{-1})
\end{multline*}
Without loss of generality, suppose that $\mathbf{Z}_a(\vtor{\hat\theta})$ is ordered so that $\hat\theta_1=...=\hat\theta_r=0$, where $r=M-\|\vtor{\hat\theta}\|_0$ and $\hat\theta_m \neq 0$ for $m > r$. Let $\mathbf{K}\triangleq \mathbf{P}\mathbf{G}(\mathbf{H})\mathbf{P}^\prime$. Then, $[\mathbf{K}+\mathbf{Z}_a(\vtor{\hat\theta})/2]^{-1}$ equals
%
%
\begin{equation*}
	\left( \begin{array}{cc}
	        \mathbf{F}_{11}^{-1} & -\widetilde{\mathbf{K}}_{11}^{-1}\mathbf{K}_{12}\mathbf{F}_{22}^{-1} \\
		-\mathbf{F}_{22}^{-1}\mathbf{K}_{21}\widetilde{\mathbf{K}}_{11}^{-1} & \mathbf{F}_{22}^{-1} 
	       \end{array}
	\right)
\end{equation*}
where $\widetilde{\mathbf{K}}_{11}=\mathbf{K}_{11}+\mathbf{Z}_a(\vtor{\hat\theta})_{11}$, $\widetilde{\mathbf{K}}_{22}=\mathbf{K}_{22}+\mathbf{Z}_a(\vtor{\hat\theta})_{22}$, $\mathbf{F}_{11}=\widetilde{\mathbf{K}}_{11}-\mathbf{K}_{12}\widetilde{\mathbf{K}}_{22}^{-1}\mathbf{K}_{21}$, and $\mathbf{F}_{22}=\widetilde{\mathbf{K}}_{22}-\mathbf{K}_{21}\widetilde{\mathbf{K}}_{11}^{-1}\mathbf{K}_{12}$. $\widetilde{\mathbf{K}}_{11}$ is invertible for sufficiently small $a$. Likewise, for sufficiently small $a$, $\widetilde{\mathbf{K}}_{22}$ is invertible by Prop.~\ref{prop:v_1}.

As $a\rightarrow 0$, $\widetilde{\mathbf{K}}_{11}^{-1}\rightarrow \mathbf{0}$ and $\widetilde{\mathbf{K}}_{22}\rightarrow \mathbf{G}(\mathbf{H})_{22}$. In addition, $\mathbf{F}_{11}^{-1}\rightarrow\mathbf{0}$ and $\mathbf{F}_{22}\rightarrow\widetilde{\mathbf{K}}_{22}$. So
%
%
\begin{equation} \label{eqn:sure1:asym}
 \mathbf{K}\left[\mathbf{K}+\frac{\mathbf{Z}_a(\vtor{\hat\theta})}{2}\right]^{-1} \rightarrow
	\left( \begin{array}{cc}
		\mathbf{0} & \mathbf{K}_{12}\mathbf{K}_{22}^{-1} \\
		\mathbf{0} & \mathbf{I}_{22}
		\end{array}
	\right)
\end{equation}
as $a\rightarrow 0$. Consequently,
%
%
\begin{equation} \label{eqn:sure1:asym2}
\lim_{a\rightarrow 0} \hat{R}_l(\vtor\zeta;a) = 
	\sigma^2 + \frac{1}{N}\|\vtor e\|_2^2 +
	\frac{2\sigma^2}{N}\|\hat{\vtor\theta}\|_0 = \hat{R}_l(\vtor\zeta).
\end{equation}

\subsection{Proof of Thm.~\ref{thm:sure2}}

Earlier notation from this appendix will be retained. The proof of the following proposition is omitted due to a lack of space.
%
%
\begin{prop} \label{prop:v_2}
 Suppose that $\mathbf{H}$ has linearly independent columns. If $\text{det}(\mathbf{P}[\mathbf{G}(\mathbf{H})-\frac{1}{2}\mathbf{U}(\vtor{\hat\theta})]\mathbf{P}^\prime)=0$, then $\mathbf{G}(\mathbf{H})$ has an eigenvalue of $1/2$.
\end{prop}
The proof of Thm.~\ref{thm:sure2} parallels the proof of Thm.~\ref{thm:sure1}. As $\Psi_{\vtor\zeta,\textit{hy}}(\vtor\theta)$ is not twice differentiable on $\mathbf{R}^M$, consider instead
%
%
\begin{multline} 
 \Psi_{\vtor\zeta,\textit{hy}}(\vtor\theta;a) = \Psi_{\vtor\zeta,l}(\vtor\theta;a) \\
+ \sum_{m=1}^M [G_1(\theta_m-\Delta_\zeta;a) - G_1(\theta_m+\Delta_\zeta;a)]
	\label{eqn:sure2:psi_relax} 
\end{multline}
where
\begin{equation}
G_1(x;a) \triangleq
	\frac{(a^2 + x^2)\arctan(x/a)-ax}{2\pi} + \frac{1}{4}x^2
	\label{eqn:sure2:G1}.
\end{equation}
$\Psi_{\vtor\zeta,\textit{hy}}(\vtor\theta;a)$ is twice differentiable in $\mathbf{R}^M$ and $\lim_{a\rightarrow 0} \Psi_{\vtor\zeta,\textit{hy}}(\vtor\theta;a) = \Psi_{\vtor\zeta,\textit{hy}}(\vtor\theta)$ pointwise. Result (\ref{eqn:sure1:rhat_inverse_prob}) can be applied to get
%
%
\begin{multline} \label{eqn:sure2:rhat_relax}
 \hat{R}_\textit{hy}(\vtor\zeta;a)=\sigma^2+\frac{\|\vtor e\|_2^2}{N} \\
	+\frac{2\sigma^2}{N}\text{tr}(
		\mathbf{G}(\mathbf{H})[\mathbf{G}(\mathbf{H})+
			\frac{1}{2}\mathbf{Z}_a(\vtor{\hat\theta})-
			\frac{1}{2}\mathbf{U}_a(\vtor{\hat\theta})]^{-1})
\end{multline}
with
%
%
\begin{equation} \label{eqn:sure2:Ua}
 \mathbf{U}_a(\vtor\theta) \triangleq \text{diag}((
	\ddot{G}_1(\theta_m-\Delta_\zeta;a)-
	\ddot{G}_1(\theta_m+\Delta_\zeta;a))_{m=1}^M)
\end{equation}
Notice that similarity between $\Psi_{\vtor\zeta,l}(\vtor\theta;a)$ and $\Psi_{\vtor\zeta,\textit{hy}}(\vtor\theta;a)$; the same applies to $\hat{R}_l(\vtor\zeta;a)$ and $\hat{R}_\textit{hy}(\vtor\zeta;a)$. 
The steps of Thm.~\ref{thm:sure1} can be carried out to evaluate the $\text{tr}(\cdot)$ expression in (\ref{eqn:sure2:rhat_relax}) as $a \rightarrow 0$. One arrives at 
%
%
\begin{equation} \label{eqn:sure2:trace}
\lim_{a \rightarrow 0} \text{tr} \left(
	\mathbf{K}_{22}[\mathbf{K}_{22}-\frac{1}{2}(\mathbf{P}\mathbf{U}_a(\vtor{\hat\theta}))_{22}]^{-1}
\right).
\end{equation}
Now $\lim_{a \rightarrow 0} \mathbf{U}_a(\vtor{\hat\theta})=\mathbf{U}(\vtor{\hat\theta})$. By assumption, $\mathbf{G}(\mathbf{H})$ does not have an eigenvalue of $1/2$. Therefore, application of Prop.~\ref{prop:v_2} implies that the inverse in (\ref{eqn:sure2:trace}) exists.

%% file: d7_double.bbl
\begin{thebibliography}{10}
\providecommand{\url}[1]{#1}
\csname url@rmstyle\endcsname
\providecommand{\newblock}{\relax}
\providecommand{\bibinfo}[2]{#2}
\providecommand\BIBentrySTDinterwordspacing{\spaceskip=0pt\relax}
\providecommand\BIBentryALTinterwordstretchfactor{4}
\providecommand\BIBentryALTinterwordspacing{\spaceskip=\fontdimen2\font plus
\BIBentryALTinterwordstretchfactor\fontdimen3\font minus
  \fontdimen4\font\relax}
\providecommand\BIBforeignlanguage[2]{{%
\expandafter\ifx\csname l@#1\endcsname\relax
\typeout{** WARNING: IEEEtran.bst: No hyphenation pattern has been}%
\typeout{** loaded for the language `#1'. Using the pattern for}%
\typeout{** the default language instead.}%
\else
\language=\csname l@#1\endcsname
\fi
#2}}

\bibitem{modis04}
Y.~{M}odis, S.~{O}gata, D.~{C}lements, and S.~C. {H}arrison, ``Structure of the
  dengue virus envelope protein after membrane fusion,'' \emph{Nature}, vol.
  427, pp. 313--319, 2004.

\bibitem{muller99}
D.~J. {M}\"uller, D.~{F}otiadis, S.~{S}cheuring, S.~A. {M}\"uller, and
  A.~{E}ngel, ``Electrostatically balanced subnanometer imaging of biological
  specimens by atomic force microscope,'' \emph{Biophysical {J}ournal},
  vol.~76, pp. 1101--1111, 1999.

\bibitem{kuznetsov01}
Y.~G. {K}uznetsov, A.~J. {M}alkin, R.~W. {L}ucas, M.~{P}lomp, and
  A.~{M}c{P}herson, ``Imaging of virus by atomic force microscopy,''
  \emph{Journal of {G}eneral {V}irology}, vol.~82, pp. 2025--2034, 2001.

\bibitem{rugar04nature}
D.~{R}ugar, R.~{B}udakian, H.~J. {M}amin, and B.~W. {C}hui, ``{Single spin
  detection by magnetic resonance force microscopy},'' \emph{Nature}, vol. 430,
  no. 6997, pp. 329--332, 2004.

\bibitem{degen08_submitted}
C.~L. {D}egen, M.~{P}oggio, H.~J. {M}amin, C.~T. {R}ettner, and D.~{R}ugar,
  ``Magnetic resonance imaging of a biological sample with nanometer
  resolution,'' \emph{Science}, submitted.

\bibitem{markiewicz95}
P.~{M}arkiewicz and M.~C. {G}oh, ``Atomic force microscope tip deconvolution
  using calibration arrays,'' \emph{Rev.~{S}ci.~{I}nstrum.}, vol.~66, pp.
  3186--3190, 1995.

\bibitem{tropp04}
J.~A. {T}ropp, ``{G}reed is good: algorithmic results for sparse
  approximation,'' \emph{{IEEE} Trans. Inform. Theory}, vol.~50, no.~10, pp.
  2231--2241, 2004.

\bibitem{fuchs05}
J.~J. {F}uchs, ``{R}ecovery of exact sparse representations in the presence of
  bounded noise,'' \emph{{IEEE} Trans. Inform. Theory}, vol.~51, no.~10, pp.
  3601--3608, 2005.

\bibitem{donoho06}
D.~L. {D}onoho, M.~{E}lad, and V.~N. {T}emlyakov, ``{S}table recovery of sparse
  overcomplete representations in the presence of noise,'' \emph{{IEEE} Trans.
  Inform. Theory}, vol.~52, no.~1, pp. 6--18, 2006.

\bibitem{tropp06}
J.~A. {T}ropp, ``{J}ust {R}elax: convex programming methods for identifying
  sparse signals in noise,'' \emph{{IEEE} Trans. Inform. Theory}, vol.~52,
  no.~3, pp. 1030--1051, 2006.

\bibitem{donoho06_stomp}
D.~L. {D}onoho, Y.~{T}saig, I.~{D}rori, and J.-L. {S}tarck, ``{S}parse solution
  of underdetermined linear equations by stagewise orthogonal matching
  pursuit,'' {S}tanford {U}niversity, Tech. Rep., 2006.

\bibitem{tibshirani96}
R.~{T}ibshirani, ``{R}egression shrinkage and selection via the lasso,''
  \emph{{J}ournal of the {R}oyal {S}tatistical {S}ociety, {S}eries {B}},
  vol.~58, no.~1, pp. 267--288, 1996.

\bibitem{alliney94}
S.~{A}lliney and S.~A. {R}uzinsky, ``{A}n algorithm for the minimization of
  mixed $l_1$ and $l_2$ norms with application to {B}ayesian estimation,''
  \emph{{IEEE} Trans. Signal Processing}, vol.~42, no.~3, pp. 618--627, 1994.

\bibitem{wipf04}
D.~P. {W}ipf and B.~D. {R}ao, ``{S}parse {B}ayesian learning for basis
  selection,'' \emph{{IEEE} Trans. Signal Processing}, vol.~52, no.~8, pp.
  2153--2164, 2004.

\bibitem{johnstone04}
I.~M. {J}ohnstone and B.~W. {S}ilverman, ``{N}eedles and straw in haystacks:
  empirical {B}ayes estimates of possibly sparse sequences,'' \emph{{T}he
  {A}nnals of {S}tatistics}, vol.~32, no.~4, pp. 1594--1649, 2004.

\bibitem{figueiredo03}
M.~A.~T. {F}igueiredo and R.~D. {N}owak, ``{An EM Algorithm for Wavelet-Based
  Image Restoration},'' \emph{{IEEE} Trans. Image Processing}, vol.~12, no.~8,
  pp. 906--916, 2003.

\bibitem{daubechies04}
I.~{D}aubechies, M.~{D}efrise, and C.~{de Mol}, ``{An Iterative Thresholding
  Algorithm for Linear Inverse Problems with a Sparsity Constraint},''
  \emph{{Communications on Pure and Applied Mathematics}}, vol.~57, no.~11, pp.
  1413--1457, 2004.

\bibitem{donoho95}
D.~L. {D}onoho and I.~M. {J}ohnstone, ``{A}dapting to unknown smoothness via
  wavelet shrinkage,'' \emph{{J}ournal of the {A}merican {S}tatistical
  {A}ssociation}, vol.~90, no. 423, pp. 1200--1224, 1995.

\bibitem{ng99}
L.~{N}g and V.~{S}olo, ``{O}ptical flow estimation using adaptive wavelet
  zeroing,'' in \emph{{P}roceedings of the {IEEE} {I}ntl.~{C}onf.~on {I}mage
  {P}rocessing}, vol.~3, 1999, pp. 722--726.

\bibitem{yuan05}
M.~{Y}uan and Y.~{L}in, ``{E}fficient empirical {B}ayes variable selection and
  estimation in linear models,'' \emph{{J}ournal of the {A}merican
  {S}tatistical {A}ssociation}, vol. 100, no. 472, pp. 1215--1225, 2005.

\bibitem{thompson91}
A.~M. {T}hompson, J.~C. {B}rown, J.~W. {K}ay, and D.~M. {T}itterington, ``{A}
  study of methods of choosing the smoothing parameter in image restoration by
  regularization,'' \emph{{IEEE} Trans. Pattern Anal. Machine Intell.},
  vol.~13, no.~4, pp. 326--339, 1991.

\bibitem{fesslerbook}
J.~A. {F}essler, ``{Image Reconstruction: Algorithms and Analysis},'' draft of
  book.

\bibitem{stein81}
C.~M. {S}tein, ``{E}stimation of the mean of a multivariate normal
  distribution,'' \emph{{T}he {A}nnals of {S}tatistics}, vol.~9, no.~6, pp.
  1135--1151, 1981.

\bibitem{byrne04}
C.~{B}yrne, ``{A unified treatment of some iterative algorithms in signal
  processing and image reconstruction},'' \emph{{Inverse Problems}}, vol.~20,
  no.~1, pp. 103--120, 2004.

\bibitem{candes05}
E.~{C}andes, J.~{R}omberg, and T.~{T}ao, ``{S}table signal recovery from
  incomplete and inaccurate measurements,'' \emph{Comm.~{P}ure {A}ppl.~{M}ath},
  vol.~59, pp. 1207--1223, 2005.

\bibitem{herrity06}
K.~K. {H}errity, A.~C. {G}ilbert, and J.~A. {T}ropp, ``{S}parse approximation
  via iterative thresholding,'' in \emph{{P}roceedings of the {IEEE}
  {I}ntl.~{C}onf.~on {A}coustics, {S}peech, and {S}ignal {P}rocessing}, 2006.

\bibitem{tingthesis}
M.~Y.~J. {T}ing, ``{S}ignal processing for magnetic resonance force
  microscopy,'' Ph.D. dissertation, {T}he {U}niversity of {M}ichigan, 2006.

\bibitem{chou94newer}
S.~H. {C}hou, L.~{Z}hu, and B.~R. {R}edi, ``{T}he {U}nusual {S}tructure of the
  {H}uman {C}entromere ({GGA})2 {M}otif: {U}npaired {G}uanosine {R}esidues
  {S}tacked {B}etween {S}heared {G}$\cdot${A} {P}airs,'' \emph{{J}.~{M}olecular
  {B}iology}, vol. 244, no.~3, pp. 259--268, 1994.

\bibitem{efron01}
B.~{E}fron, ``{Selection criteria for scatterplot smoothers},'' \emph{{The
  Annals of Statistics}}, vol.~29, no.~2, pp. 470--504, 2001.

\bibitem{mackay99}
D.~J.~C. {M}ackay, ``{C}omparison of {A}pproximate {M}ethods for {H}andling
  {H}yperparameters,'' \emph{{N}eural {C}omputation}, vol.~11, pp. 1035--1068,
  1999.

\bibitem{lange00}
K.~{L}ange, D.~R. {H}unter, and I.~{Y}ang, ``{O}ptimization transfer using
  surrogate objective functions,'' \emph{{J}ournal of {C}omputational and
  {G}raphical {S}tatistics}, vol.~9, no.~1, pp. 1--20, 2000.

\bibitem{solo96}
V.~{S}olo, ``A sure-fired way to choose smoothing parameters in ill-conditioned
  inverse problems,'' in \emph{{P}roceedings of the {IEEE} {I}ntl.~{C}onf.~on
  {I}mage {P}rocessing}, vol.~3, 1996, pp. 89--92.

\end{thebibliography}
